\documentclass[a4paper,11pt]{article}

\usepackage{fullpage}
\usepackage[T1]{fontenc}
\usepackage{xcolor,cite,hyperref,caption}
\usepackage{amssymb,amsmath,mathtools,slashed,mathrsfs}
\usepackage{graphicx,ulem}
\usepackage{braket}

\newcommand{\be}{\begin{eqnarray}}
\newcommand{\ee}{\end{eqnarray}}

\allowdisplaybreaks

\numberwithin{equation}{section}

\begin{document}

\setcounter{footnote}{0}

\baselineskip 6 mm

\begin{titlepage}
	\thispagestyle{empty}
	\begin{flushright}
		
	\end{flushright}

	\vspace{35pt}
	
	\begin{center}
	    { \LARGE{\bf Scale separation and partial supersymmetry breaking}} 
		
		\vspace{50pt}
		
		{\Large Niccol\`o Cribiori$^1$,  Fotis Farakos$^2$,  Alexandros Zarafonitis$^2$}  
		
		\vspace{25pt}

        $^1${\it KU Leuven, Institute for Theoretical Physics, \\ 
        Celestijnenlaan 200D, B-3001 Leuven, Belgium}
		
		\vspace{15pt}

        $^2${\it Physics Division, National Technical University of Athens \\
        15780 Zografou Campus, Athens, Greece}
		
		\vspace{15pt}

		\vspace{40pt}
		
		{ABSTRACT} 
	\end{center}

We propose new candidate string theory vacua in which partial supersymmetry breaking occurs within the lower-dimensional effective theory in anti-de Sitter space, parametrically below the Kaluza--Klein scale. These vacua solve the equations of motion of massless type IIA supergravity in the smeared approximation and exhibit parametric scale separation, at both weak and strong string coupling. They are reproduced as $\mathcal N=1$ supersymmetric phases of a three-dimensional $\mathcal N=2$ gauged supergravity. Should these vacua be solutions of string theory, they would provide explicit setups in which partial supersymmetry breaking can be studied at energies parametrically below the compactification scale. In addition, we further comment on our previously proposed solution with scale separation and extended supersymmetry.

\vspace{10pt}

\bigskip

\end{titlepage}

\newpage

\tableofcontents

\newpage

\section{Introduction}

Two of the most obvious properties of the universe from the point of view of a low-energy observer are that it appears four-dimensional and non-supersymmetric. Explaining either property microscopically is a central open problem, especially in string theory. Whenever these properties are linked for fundamental reasons, so that neither can be studied separately, the challenge becomes even more severe.

Solutions of string theory in which the ten-dimensional spacetime contains a compact internal factor with a characteristic length much smaller than that of the non-compact external dimensions are called scale-separated \cite{Coudarchet:2023mfs}. They are notoriously difficult to construct: even at the level of the effective supergravity description, the main candidate for many years was \cite{DeWolfe:2005uu,Camara:2005dc}, until \cite{Farakos:2020phe,Cribiori:2021djm} put forward new models. Since then, several scale-separated supergravity vacua have been proposed, many featuring a three-dimensional anti-de Sitter external space; see, e.g., \cite{Emelin:2021gzx,VanHemelryck:2022ynr,Farakos:2023nms,Farakos:2023wps,Arboleya:2024vnp,VanHemelryck:2025qok,Miao:2025rgf,Tringas:2025bwe,Arboleya:2025jko,Cribiori:2026caf,Aggarwal:2026jan}. These setups preserve at most minimal supersymmetry. Our very recent construction \cite{Cribiori:2026caf} was proposed as an exception, realizing scale separation together with extended supersymmetry. In this work, we clarify the status of that claim by reconsidering the supersymmetry preserved by the full source configuration.
Indeed, bottom-up arguments suggest that scale separation with extended supersymmetry might be difficult to realize, or perhaps even impossible \cite{Cribiori:2022trc,Montero:2022ghl,Cribiori:2023ihv,Cribiori:2024jwq}. The three-dimensional case is an exception to the scope of these arguments, making it particularly interesting to investigate. Also worth noting is the very recent work \cite{Tringas:2026ptt} in two dimensions, which may constitute a counterexample to the no-go result of \cite{Cribiori:2024jwq}, depending on the assumptions entering the effective theory cutoff.\footnote{The question is whether the ultraviolet cutoff must be bounded from above by the scale set by the tension of a BPS domain wall, as, e.g., in \cite{Lanza:2020qmt,Martucci:2022krl,Martucci:2024trp}, even if this domain wall is heavy and thus unlikely to be produced. While we do not have a definitive answer, we would like to point out an analogy with ten-dimensional supergravity, whose effective description is valid only up to the string scale even when string excitations are unlikely to be produced. Neglecting this limitation, one might be tempted to declare the theory valid up to the ten-dimensional Planck scale, which is parametrically higher than the string scale at weak coupling.}

Supersymmetry breaking, on the other hand, is ubiquitous in string theory. Fundamental dynamical objects of the theory, such as D-branes, can spontaneously break part of the bulk supersymmetry, with the broken supersymmetries subsequently realized nonlinearly. However, despite the presence of several supersymmetry breaking ingredients, constructing controlled vacua with broken supersymmetry has proved challenging. A central problem is how to justify an effective spacetime description when there is insufficient supersymmetry to control corrections to the leading-order effective action. One may hope that, if supersymmetry breaking is only partial, some of this control can be retained. Nevertheless, partial supersymmetry breaking is a notoriously intricate mechanism to realize \cite{Cecotti:1984rk,Antoniadis:1995vb}, and explicit constructions are rare.

In the simplest cases, the supersymmetry-breaking scale in string theory is set by the string scale, as in brane supersymmetry breaking, or by the Kaluza--Klein scale, as in Scherk--Schwarz compactifications. Obtaining supersymmetry breaking at even lower scales in explicit and controlled models is more difficult. Examples of partial breaking have been proposed in Minkowski \cite{Frey:2002hf,Tripathy:2002qw,Andrianopoli:2003jf}, while partial supersymmetry breaking in anti-de Sitter has been studied in \cite{House:2005yc,Louis:2009xd}. However, Minkowski vacua are automatically scale-separated, so they shed little light on the dynamical origin of small extra dimensions. Conversely, the aforementioned anti-de Sitter constructions do not establish partial supersymmetry breaking together with parametric scale separation. Nonsupersymmetric branches of known scale separated solutions, such as \cite{DeWolfe:2005uu}, do exhibit a gravitino mass parametrically below the Kaluza-Klein scale, but the breaking is complete rather than partial, and so one has to worry about all possible corrections to the spacetime effective action.

In this work, we propose a construction in which partial supersymmetry breaking and scale separation are both consequences of the vacuum dynamics. Our model gives three-dimensional $\mathcal N=1$ anti-de Sitter solutions of the ten-dimensional massless type IIA supergravity with smeared O6-planes, featuring parametric scale separation. All internal radii can become parametrically large in string units, both at weak and at strong string coupling. These solutions are also reproduced as supersymmetric vacua of a three-dimensional $\mathcal N=2$ gauged supergravity, in which one supersymmetry is spontaneously broken at the anti-de Sitter scale. The spontaneous breaking of supersymmetry therefore occurs parametrically below the Kaluza--Klein scale.

The interplay between the ten-dimensional solution and its effective description is nontrivial. We first establish an obstruction to strict $G_2$ supersymmetric solutions in massless type IIA within the ansatz considered here, motivating the construction of solutions with $\mathrm{SU}(3)$ structure, which we determine explicitly. We then exclude an $\mathcal N=(1,1)$ enhancement of such solutions compatible with our orbifold and orientifold projections. Nevertheless, we notice that a putative holomorphic superpotential reproduces the three-dimensional scalar potential at vanishing axions and appears to predict precisely such an enhancement. We resolve this apparent contradiction by performing the direct dimensional reduction and identifying axionic gaugings that the superpotential description fails to capture. Including these gaugings reproduces the correct effective action and gives an $\mathcal N=1$ anti-de Sitter vacuum of the three-dimensional $\mathcal N=2$ gauged supergravity theory. 

We also revisit our previous construction \cite{Cribiori:2026caf}. The extended supersymmetry identified there concerns the proposed effective description of the untwisted closed string sector in the smeared approximation. The combined O6/D6 projections instead indicate that the full source configuration may preserve only minimal supersymmetry in the toroidal orbifold limit. Since D-branes do not impose an orientifold truncation, we expect the additional supersymmetry to be spontaneously broken and nonlinearly realized, although establishing this interpretation requires including explicitly the open string sector in the effective theory.

These results suggest that, even when extended supersymmetry seems naively possible, scale separation may favor minimal supersymmetry through a nontrivial interplay between partial breaking and restrictions imposed by the geometry and sources. Our relatively simple setup provides a valuable starting point for investigating the relation between supersymmetry breaking and the smallness of the extra dimensions, while still preserving minimal supersymmetry.

\section{Model}
\label{sec:themodel}

The model we will be considering arises as a variant of the setup recently put forward in \cite{Tringas:2026ncg}, so that we will refer to that work for notation and conventions and we will recall only the key aspects of the constructions which are necessary to understand the difference with our new setup. Indeed, despite the formal similarity with \cite{Tringas:2026ncg} at the level of the microscopic setup, the  dynamics of the lower dimensional effective theory will contain essential differences.  We will work in string units $2\pi\sqrt{\alpha'}=1$.

\subsection{Geometry of the internal space}
\label{sec:geometry}

We consider massless type IIA supergravity compactified on an orientifold of
the toroidal orbifold $\tilde T^6/\mathbb{Z}_2\times\mathbb{Z}_2$, where
$\tilde T^6$ is a twisted six-torus. We further compactify on a circle,
without introducing any additional orbifold action, so that the internal
seven-dimensional space becomes
\begin{equation}
    \mathcal{M}_7=\frac{\tilde T^6}{\mathbb{Z}_2\times \mathbb{Z}_2}\times S^1\,.
    \label{eq:M7}
\end{equation}
This setup is a variant of the one recently put forward in \cite{Tringas:2026ncg}. Locally, the two geometries coincide, but we do not impose the additional $\widehat{\mathbb{Z}}_2$ orbifold of \cite{Tringas:2026ncg}, which reduces the supersymmetry of the three-dimensional theory from $\mathcal{N}=2$ to $\mathcal{N}=1$. The orbifold and orientifold projections we retain are instead compatible with $\mathcal{N}=2$ supersymmetry in three dimensions. Hence, the fact that our vacua preserve only $\mathcal{N}=1$ cannot be attributed to any of the projections defining the compactification, but will be a consequence of the dynamics.

The six-dimensional space $\tilde T^6$ is the Iwasawa nilmanifold, which appeared as a type IIA background already in \cite{Lust:2004ig,Caviezel:2008ik,Koerber:2008rx} and underlies the scale-separated vacua of \cite{Cribiori:2021djm}. It is conveniently described in terms of a basis of left-invariant one-forms $\eta^i$, with
$i=1,2,3,5,6,7$ (following the conventions of \cite{Tringas:2026ncg}, the label $i=4$ is reserved for the additional circle), satisfying the Maurer--Cartan equations
\begin{equation}
    d\eta^a=-\frac12 f^a{}_{bc}\,\eta^b\wedge\eta^c\,,
\end{equation}
where the only non-vanishing structure constants, also called metric fluxes, are
\begin{equation}
    f^1{}_{27}=f^1{}_{36}=-h\,,\qquad f^5{}_{23}=f^5{}_{67}=h\,.
\end{equation}
Explicitly, this gives
\begin{equation}
    d\eta^1=h\left(\eta^{27}+\eta^{36}\right),\qquad
    d\eta^5=-h\left(\eta^{23}+\eta^{67}\right),\qquad
    d\eta^i=0\,,\quad i\neq 1,5\,,
    \label{eq:MC}
\end{equation}
where we use the shorthand notation $\eta^{ij}=\eta^i\wedge\eta^j$.
The geometry of $\tilde T^6$ can be visualized as follows. The closed one-forms $\eta^2,\eta^3,\eta^6,\eta^7$ span a base $T^4$, while the directions along $\eta^1$ and $\eta^5$ are two circle fibers: the $\eta^1$ circle is twisted over the two-tori in the base spanned by the $(\eta^2,\eta^7)$ and $(\eta^3,\eta^6)$ directions, while the $\eta^5$ circle is twisted over those spanned by the $(\eta^2,\eta^3)$ and $(\eta^6,\eta^7)$ directions. Hence, $\tilde T^6$ is a $T^2$ bundle over $T^4$. The curvatures of the two circle fibrations are $d\eta^1$ and $d\eta^5$ in \eqref{eq:MC}, which represent their first Chern classes, and their periods over these two-tori are $\pm h$. Since all circles are normalized to have unit length, integrality of the Chern classes requires $h\in\mathbb{Z}$.

We introduce the vielbein one-forms $e^i=r_i\,\eta^i$, with no sum over
$i=1,\dots,7$, where the $r_i$ are the radii in string units, so that the
internal seven-dimensional metric reads
\begin{equation}
    ds^2_7=\sum_{i=1}^7\left(e^i\right)^2.
\end{equation}
The additional circle is untwisted, and the corresponding unit one-form,
$v\equiv e^4$, is closed, 
\begin{equation}
dv=0. 
\end{equation}
It will play a distinguished role in what follows.

Recall that an $\mathrm{SU}(3)$ structure on a six-dimensional manifold is specified by a real two-form $J$ and a locally decomposable complex three-form $\Omega$, satisfying
\begin{equation}
    J\wedge\Omega=0\,,\qquad
    \frac{i}{8}\,\Omega\wedge\bar\Omega=\frac16 J\wedge J\wedge J=\mathrm{vol}_6\,,
\end{equation}
with ${\rm vol}_6$ the volume form. Besides, in our conventions $*_6 J = \frac12 J \wedge J$ and $*_6 \Omega =-  i \Omega$. 
Since $\mathcal{M}_7$ is a product, these forms, together with $v$, define an $\mathrm{SU}(3)$ structure on $\mathcal{M}_7$: the subgroup of $\mathrm{SO}(7)$ preserving $v$, $J$ and $\Omega$ is the $\mathrm{SU}(3)$ acting on the directions orthogonal to $v$. On a generic $\mathrm{SU}(3)$-structure manifold, the exterior derivatives of $J$ and $\Omega$ decompose into five torsion classes \cite{chiossi2002,LopesCardoso:2002vpf}. In our case, the $\mathrm{SU}(3)$ structure is half-flat, namely
\begin{equation}
    d\,\mathrm{Im}\,\Omega=0\,,\qquad d\left(J\wedge J\right)=0\,,
    \label{eq:half-flat}
\end{equation}
and only three torsion classes survive,
\begin{equation}
    dJ=\frac32\, w_1\,\mathrm{Im}\,\Omega+w_3\,,\qquad
    d\,\mathrm{Re}\,\Omega=w_1\, J\wedge J+w_2\wedge J\,.
    \label{eq:torsion-classes}
\end{equation}
Here, $w_1$ is a real function, $w_2$ is a real primitive $(1,1)$-form, $w_2\wedge J\wedge J=0$, $*_6 (w_2 \wedge J)=-w_2$, and $w_3$ is a real primitive $((2,1)+(1,2))$-form, $w_3\wedge J=w_3\wedge\Omega=0$.
Half-flat structures of this type appear in type IIA anti-de Sitter vacua with O6-planes \cite{Lust:2004ig,Caviezel:2008ik}. 
Since $w_1$ and $w_2$ measure the obstruction to the integrability of the almost complex structure \cite{LopesCardoso:2002vpf}, the latter is non-integrable in our case, and $\Omega$ is a $(3,0)$-form but not a holomorphic one.

\subsection{Orbifold and O6-planes}

We now introduce the orbifold and orientifold projections compatible with the $\mathrm{SU}(3)$-structure geometry just described. We consider four O6-planes, each filling the external three-dimensional spacetime and wrapping a four-dimensional internal cycle, given by the fixed locus of the corresponding orientifold involution. The configuration is summarized in table~\ref{tab:O6} and coincides with that of \cite{Tringas:2026ncg}, Figure 1.
\begin{table}[h]
\centering
\begin{tabular}{|c|c|c|c|c|c|c|c|c|}
\hline
& $\eta^1$ & $\eta^2$ & $\eta^3$ & $\eta^4$ & $\eta^5$ & $\eta^6$ & $\eta^7$
& Current \\
\hline
$O6_{\alpha}$
& $-$ & $\times$ & $\times$ & $\times$ & $\times$ & $-$ & $-$
& $e^{167}$ \\
\hline
$O6_{\alpha\beta}$
& $\times$ & $-$ & $\times$ & $\times$ & $-$ & $\times$ & $-$
& $e^{257}$ \\
\hline
$O6_{\alpha\gamma}$
& $\times$ & $\times$ & $-$ & $\times$ & $-$ & $-$ & $\times$
& $e^{356}$ \\
\hline
$O6_{\alpha\beta\gamma}$
& $-$ & $-$ & $-$ & $\times$ & $\times$ & $\times$ & $\times$
& $e^{123}$ \\
\hline
\end{tabular}
\caption{O6-plane configuration. An $\times$ denotes a direction wrapped by the O6-plane, which is invariant under the corresponding involution, while a $-$ denotes a transverse direction, which is reflected. The last column gives the smeared Poincar\'e-dual three-form current.}
\label{tab:O6}
\end{table}

The corresponding involutions act diagonally on the basis of left-invariant one-forms, flipping the sign of the directions transverse to each O6-plane. Explicitly, on $(\eta^1,\eta^2,\eta^3,\eta^4,\eta^5,\eta^6,\eta^7)$ they act as
\begin{equation}
\begin{aligned}
    \sigma_{\alpha}&:\ (-\eta^1,\eta^2,\eta^3,\eta^4,\eta^5,-\eta^6,-\eta^7)\,,\\
    \sigma_{\alpha\beta}&:\ (\eta^1,-\eta^2,\eta^3,\eta^4,-\eta^5,\eta^6,-\eta^7)\,,\\
    \sigma_{\alpha\gamma}&:\ (\eta^1,\eta^2,-\eta^3,\eta^4,-\eta^5,-\eta^6,\eta^7)\,,\\
    \sigma_{\alpha\beta\gamma}&:\ (-\eta^1,-\eta^2,-\eta^3,\eta^4,\eta^5,\eta^6,\eta^7)\,.
\end{aligned}
\end{equation}
One can check that these maps preserve the Maurer--Cartan equations \eqref{eq:MC}, and are thus well defined on $\tilde T^6$. Notice that the closed one-form $\eta^4$ is even under every involution, consistently with the fact that all four O6-planes wrap the circle.

The orbifold group $\mathbb{Z}_2\times\mathbb{Z}_2=\{1,\Theta_1,\Theta_2, \Theta_1\Theta_2\}$ acts as \cite{Tringas:2026ncg}
\begin{equation}
\begin{aligned}
    \Theta_1&=\sigma_{\alpha}\sigma_{\alpha\beta}:\ (-\eta^1,-\eta^2,\eta^3,\eta^4,-\eta^5,-\eta^6,\eta^7)\,,\\
    \Theta_2&=\sigma_\alpha \sigma_{\alpha\gamma}:\ (-\eta^1,\eta^2,-\eta^3,\eta^4,-\eta^5,\eta^6,-\eta^7)\,,\\
    \Theta_1\Theta_2&= \sigma_{\alpha\beta}\sigma_{\alpha\gamma}:\ (\eta^1,-\eta^2,-\eta^3,\eta^4,\eta^5,-\eta^6,-\eta^7)\,.
    \label{eq:OrbifoldGen}
\end{aligned}
\end{equation}
Notice that the four O6 involutions are obtained by composing the single involution $\sigma_\alpha$ with the elements of the orbifold group,
\begin{equation}
    \sigma_\alpha\Theta_1=\sigma_{\alpha\beta}\,,\qquad
    \sigma_\alpha\Theta_2=\sigma_{\alpha\gamma}\,,\qquad
    \sigma_\alpha\Theta_1\Theta_2=\sigma_{\alpha\beta\gamma}\,.
\end{equation}
Hence, the whole O6-plane configuration is generated by $\sigma_\alpha$ together with the orbifold group. Equivalently, the product of any two O6 involutions is an element of the orbifold group. This is consistent with the fact that each involution is accompanied by the worldsheet parity and $(-1)^{F_L}$, which square to one, so that the product of two orientifold actions is a purely geometric symmetry \cite{Grimm:2004ua}.

Compatibility with these projections requires the $\mathrm{SU}(3)$ structure to be invariant under the orbifold group, and to transform as
\begin{equation}
\label{eq:sigmaJOmega}
    \sigma^*J=-J\,,\qquad \sigma^*\Omega=\bar\Omega\,,
\end{equation} 
under each orientifold involution \cite{Grimm:2004ua}. The latter condition ensures that the O6-planes wrap cycles calibrated by $\mathrm{Re}\,\Omega$, and thus preserve the same supersymmetry as the background \cite{Koerber:2007jb}. These requirements restrict the forms on which $J$ and $\Omega$ can be expanded, as we discuss in the next section.

\subsection{Fluxes and Bianchi identities}
\label{sec:fluxes}

The orbifold and orientifold projections restrict the left-invariant forms
on which the internal fields can be expanded. Since $\eta^4$ is invariant
under all of them, it suffices to discuss forms on $\tilde T^6$, where one
can check that there are no invariant one-forms, while the invariant two-,
three- and four-forms are
\begin{equation}
\begin{aligned}
    \Lambda^2_- (\tilde T^6)&=\mathrm{span}\left(\eta^{15},\eta^{26},\eta^{37}\right),
    &\Lambda^2_+ (\tilde T^6) &=0\,,\\
    \Lambda^3_- (\tilde T^6) &=\mathrm{span}\left(\eta^{123},\eta^{167},\eta^{257},\eta^{356}\right),
    &\Lambda^3_+ (\tilde T^6) &=\mathrm{span}\left(\eta^{127},\eta^{136},\eta^{235},\eta^{567}\right),\\
    \Lambda^4_+ (\tilde T^6) &=\mathrm{span}\left(\eta^{1256},\eta^{1357},\eta^{2367}\right),
    &\Lambda^4_- (\tilde T^6) &=0\,.
\end{aligned}
\label{eq:invariant-forms}
\end{equation}
Here, the $\pm$ subscript denotes the parity under the orientifold involution, and since all four $\sigma$'s differ by orbifold elements, they act in the same way on orbifold-invariant forms. In particular, the volume form is odd. These projections also fix the form of the metric. The only pairs of one-forms with the same orbifold parities are $(\eta^1,\eta^5)$, $(\eta^2,\eta^6)$ and $(\eta^3,\eta^7)$, which however have opposite parity under $\sigma$. Hence, no off-diagonal terms are allowed, and the diagonal metric introduced in section~\ref{sec:geometry} is the most general left-invariant metric compatible with the projections.

The $\mathrm{SU}(3)$ structure must satisfy the requirements \eqref{eq:sigmaJOmega}, namely $J\in\Lambda^2_- (\tilde T^6)$, $\mathrm{Re}\,\Omega\in\Lambda^3_+ (\tilde T^6)$ and
$\mathrm{Im}\,\Omega\in\Lambda^3_- (\tilde T^6)$. We take \cite{Tringas:2026ncg}
\begin{align}
    J&=r_3r_7\,\eta^{37}-r_1r_5\,\eta^{15}-r_2r_6\,\eta^{26}\,,
    \label{Kahler_expansion} \\
    \mathrm{Re}\,\Omega&=r_1r_2r_7\,\eta^{127} +r_1r_3r_6\,\eta^{136}-r_2r_3r_5\,\eta^{235}-r_5r_6r_7\,\eta^{567}\,, \label{ReΩ_expansion} \\
    \mathrm{Im}\,\Omega&=r_2r_5r_7\,\eta^{257}+r_3r_5r_6\,\eta^{356}-r_1r_2r_3\,\eta^{123}-r_1r_6r_7\,\eta^{167}\,, \label{ImΩ_expansion}
\end{align}
which corresponds to $\Omega=i\,(e^5+ie^1)\wedge(e^6+ie^2)\wedge(e^3+ie^7)$ and gives $\mathrm{vol}_6=-e^{123567}$.

We now identify the scalar fields of the lower-dimensional theory. Since $\tilde T^6$ is a nilmanifold, we can truncate to the left-invariant modes compatible with the orbifold and orientifold projections \cite{Hull:2005hk,Caviezel:2008ik}, which are thus counted by the invariant forms \eqref{eq:invariant-forms}. In the metric sector, the $\mathrm{SU}(3)$ structure is specified by $J\in\Lambda^2_-(\tilde T^6)$, which contains three real parameters, and by $\mathrm{Re}\,\Omega\in\Lambda^3_+(\tilde T^6)$, which contains four real parameters, of which one is fixed in terms of $J$ by the compatibility conditions. These six parameters are modeled by the scalars
\begin{equation}
    t_1=r_1r_5\,,\qquad t_2=r_2r_6\,,\qquad t_3=r_3r_7\,,\qquad
    u_1=\frac{r_5}{r_1}\,,\qquad u_2=\frac{r_6}{r_2}\,,\qquad u_3=\frac{r_7}{r_3}\,,
    \label{eq:scalarsmodel}
\end{equation}
which in the untwisted case $h=0$, where $\tilde T^6/\mathbb{Z}_2\times \mathbb{Z}_2$ becomes the Calabi--Yau orbifold $T^6/\mathbb{Z}_2\times\mathbb{Z}_2$, are the K\"ahler and complex structure moduli of the untwisted sector \cite{Camara:2005dc}. Together with the circle radius $r_4$ and the dilaton, they are complexified by the axions arising from $B_2\in\Lambda^2_-$ and $C_3\in\Lambda^3_+$, and by the scalar dual to the Kaluza--Klein vector of the circle.

Using the Maurer--Cartan equations \eqref{eq:MC}, one finds that
$d\,\mathrm{Im}\,\Omega=0=d(J\wedge J)$, as anticipated, while
\begin{equation}
    dJ=a_1\,e^{257}+a_2\,e^{356}+a_3\,e^{167}+a_4\,e^{123}\,,\qquad
    d\,\mathrm{Re}\,\Omega=\Delta\,e^{2367}\,,
\end{equation}
where $a_i$ are the metric fluxes dressed by the radii,
\begin{equation}
\label{ai_expr}
    a_1=\frac{h\,r_1}{r_2r_7}\,,\qquad a_2=\frac{h\,r_1}{r_3r_6}\,,\qquad
    a_3=-\frac{h\,r_5}{r_6r_7}\,,\qquad a_4=-\frac{h\,r_5}{r_2r_3}\,,
\end{equation}
and $\Delta=a_1+a_2-a_3-a_4$. Comparing with \eqref{eq:torsion-classes}, we
read off the torsion classes
\begin{align}
    \label{eq:w1}
    w_1&=\frac{\Delta}{6}\,,\\
    \label{eq:w2}
    w_2&=\frac{\Delta}{3}\left(2e^{15}-e^{26}+e^{37}\right),\\
        \label{eq:w3}
    w_3&=\left(a_1-\tfrac{\Delta}{4}\right)e^{257}
    +\left(a_2-\tfrac{\Delta}{4}\right)e^{356}
    +\left(a_3+\tfrac{\Delta}{4}\right)e^{167}
    +\left(a_4+\tfrac{\Delta}{4}\right)e^{123}\,.
\end{align}

Although the truncation to left-invariant modes is consistent \cite{Hull:2005hk}, it is not by itself guaranteed to capture all light fields. The standard argument is that left-invariant modes have Laplacian eigenvalues of the order of $a_i^2$ \cite{Caviezel:2008ik}, while the higher Kaluza--Klein modes, which are truncated, have masses set by the inverse radii. Hence, the truncation captures all the light modes when
\begin{equation}
    |a_i|\,r_j\ll1\,,\qquad \forall\, i,j\,.
    \label{eq:small-flux}
\end{equation}
On the six-dimensional twisted torus, the twisting does not lower the masses of the scalar Kaluza--Klein modes with respect to the flat-torus case \cite{Cribiori:2021djm}, while for higher-degree forms its effect is negligible in the regime \eqref{eq:small-flux}, as shown explicitly for the three-dimensional Heisenberg nilmanifold in \cite{Andriot:2018tmb}. We will verify that our vacua satisfy \eqref{eq:small-flux}. A complete analysis would require the full Kaluza--Klein spectrum around the solution, which we leave for future work.

We now turn to the fluxes. We consider massless type IIA with vanishing NSNS flux, $F_0=0=H_3$. The remaining RR field strengths transform under the orientifold involution as \cite{Grimm:2004ua}
\begin{equation}
    \sigma^*F_2=-F_2\,,\qquad \sigma^*F_4=F_4\,,\qquad \sigma^*F_6=-F_6\,,
\end{equation}
so that $F_2\in\Lambda^2_-(\tilde T^6)$, $F_6$ is proportional to the volume form, and $F_4$ can have components either in $\Lambda^4_+(\tilde T^6)$ or along $\eta^4\wedge\Lambda^3_+(\tilde T^6)$. The former are forced to vanish by the Killing spinor equations, as we will see in the next section, and we set them to zero from the outset. We thus take the ansatz\footnote{The names of the flux parameters are those of \cite{Tringas:2026ncg}, inherited from the massive type IIA frame related to ours by a formal double T-duality along $\eta^1,\eta^5$, except for $f_1,\dots,f_4$, which correspond to $f_4^1,f_4^3,f_4^4,f_4^5$ there. In particular, $m_0$ descends from the Romans mass of that frame, although our setup is massless.}
\begin{align}
    F_2&=m_0\,\eta^{15}+N\,\eta^{26}-R_F\,\eta^{37}\,,\label{RR2_expansion}\\
    F_4&=\eta^4\wedge\left(f_1\,\eta^{136}-f_2\,\eta^{235}+f_3\,\eta^{127}
    -f_4\,\eta^{567}\right), \label{RR4_expansion}\\
    F_6&=-M\,\eta^{123567}\,.
    \label{RR6_expansion}
\end{align}
In the presence of O6-planes, the Bianchi identities read
\begin{equation}
    dF_2=j_{\rm O6}\,,\qquad dF_4=0\,,\qquad dF_6=0\,,
\end{equation}
where, in the smeared approximation, the O6-plane current is
\cite{Cribiori:2021djm,Tringas:2026ncg}
\begin{equation}
    j_{\rm O6}=Q_{\rm O6}\left(\eta^{257}+\eta^{356}-\eta^{167}-\eta^{123}\right),
\end{equation}
with $Q_{\rm O6}$ the O6-plane charge. Each term in the current is the Poincar\'e dual of the cycle wrapped by the corresponding set of O6-planes. The relative signs are fixed by supersymmetry: O6-planes that preserve the
same supersymmetry as the background wrap cycles calibrated by
$\mathrm{Re}\,\Omega$, and their smeared current is then proportional to
$\mathrm{Im}\,\Omega$ \cite{Koerber:2010bx}.
Since $dF_2=-m_0h\left(\eta^{257}+\eta^{356}-\eta^{167}-\eta^{123}\right)$, the tadpole cancellation condition is
\begin{equation}
   m_0 h=- Q_{\rm O6}\,, 
\end{equation}
and its solution is presented in appendix \ref{app:O6count}. Here, we just recall that we solve it by imposing $h$ even in what follows, which implies
\begin{equation}
m_0 h = 16.
\label{eq:F2tadheven}
\end{equation}
The Bianchi identity for $F_6$ is trivially satisfied, while that for $F_4$ gives the single condition
\begin{equation}
    f_1+f_2+f_3+f_4=0\,,
    \label{eq:dF4tad}
\end{equation}
since $d(\eta^{127})=d(\eta^{136})=-d(\eta^{235})=-d(\eta^{567}) =h\,\eta^{2367}$.

Finally, in units $2\pi\sqrt{\alpha'}=1$ and at the level of cohomology, flux quantization requires the periods of the RR fluxes over the cycles of the compactification space to be integers. As is common for toroidal orbifold models, see e.g.\ \cite{DeWolfe:2005uu}, we impose this condition in a natural toroidal basis of forms, here on the covering space, which gives
\begin{equation}
    N,\,R_F,\,f_1,\dots,f_4,\,M\in\mathbb{Z}\,,
\end{equation}
while $h\in2\mathbb{Z}$, as discussed in appendix~\ref{app:O6count}. On the orbifold and orientifold quotient, after resolving the orbifold singularities, these conditions may be modified by order-one factors, and further refinements may arise from K-theory, as well as from the shifted quantization of $F_4$ inherited from M-theory \cite{Diaconescu:2000wy} (the latter vanishes on the untwisted cycles, since $\tilde T^6\times S^1$ is parallelizable \cite{Koerber:2010bx}, the left-invariant one-forms $\eta^i$ providing a global frame on it). A precise determination would require the integral homology, or more precisely the K-theory, of the resolved orientifold, which is beyond the scope of this work and, to our knowledge, not available even for the benchmark model of \cite{DeWolfe:2005uu}. The key fact for us is that such factors do not affect any of the parametric statements below. They may instead restrict the allowed solutions of the tadpole condition \eqref{eq:F2tadheven}, which would however remain available even if $m_0$ is required to be a multiple of $8$, as for $(h,m_0)=(2,8)$. Notice that the parameter $m_0$ is not the period of a closed form, since $\eta^{15}$ is not closed; we take it to be an integer, as inherited from the Romans mass in the T-dual frame \cite{Tringas:2026ncg}.

\subsection{On the extended supersymmetry in our previous solution}

We would like to pause our discussion for a moment and take this opportunity to clarify an aspect of our construction in \cite{Cribiori:2026caf}. In that paper, we proposed scale-separated three-dimensional $\mathcal{N}=(1,1)$ anti-de Sitter vacua. The strategy was the following: first we solved the supersymmetry variations of ten-dimensional supergravity, ensuring the existence of at least minimal supersymmetry, and then we matched the solution thus found with the one obtained from the F-term equations of a holomorphic superpotential. Having excluded the possibility of obtaining the same solution within gauged supergravity, the latter step ensured the presence of the second supersymmetry, \textit{within the proposed effective description}. 

Crucially, the microscopic setup contains both O6-planes and D6-branes. The former are positioned in such a way as to be compatible with $\mathcal{N}=(1,1)$ supersymmetry; instead, the latter are positioned in such a way that, if they were O-planes, they would further truncate the supersymmetry by a half, as we will argue shortly. Being D-branes, however, they do not impose this orientifold truncation, and we rather expect the additional supersymmetry to remain, albeit non-linearly realized. 
We missed this effect in \cite{Cribiori:2026caf} because in that work we explicitly considered only the (untwisted) closed string moduli in the effective description. 
\textit{Hence, the statement about extended supersymmetry in \cite{Cribiori:2026caf} applies to the proposed effective description of the (untwisted) closed string sector in the smeared approximation, and does not by itself establish extended supersymmetry of the full source-coupled background.} 
While we have not performed the necessary computation, we believe that once one includes the open string sector, one will see that part of the extended supersymmetry is spontaneously broken and non-linearly realized, so that the full setup preserves only $\mathcal{N}=1$ supersymmetry at the linearly realized level and on the vacuum.
This intuition can be anticipated by looking at the projectors of the combined O6/D6-brane system, in the toroidal orbifold setup and in the smeared approximation. Away from the smeared approximation and on the smooth resolution, this simple analysis might no longer apply. However, we currently lack the means to perform the necessary computation in such regimes.

One quick way to anticipate this supersymmetry counting in the setup of \cite{Cribiori:2026caf} is to notice that the source configuration displayed in Table 1 therein formally matches that in the table in section 2.3 of \cite{Cribiori:2026qfs} upon relabelling the coordinates as $(x^1,x^2,x^3,x^4,x^5,x^6,x^7)\to \allowbreak  (e^1,e^3,e^5,e^2,e^4,e^6,e^7)$, $x^7$ being the circle direction, and mapping the sources as $(\mathrm{O6}_1,\mathrm{O6}_2,\mathrm{O6}_3,\mathrm{O6}_4, \mathrm{D6}_1,\mathrm{D6}_2,\mathrm{D6}_3) \to (\mathrm{O6}^C,\mathrm{O6}^D,\mathrm{O6}^F,\mathrm{O6}^G, \mathrm{O6}^A,\mathrm{O6}^B,\mathrm{O6}^E)$.
Since by construction the setup of \cite{Cribiori:2026qfs} preserves minimal supersymmetry, this suggests the same should happen for the combined O6/D6 system in \cite{Cribiori:2026caf}, provided sources are mutually oriented in the appropriate way.

In the toroidal orbifold limit we can verify the supersymmetry counting explicitly, following the strategy outlined in \cite{Bergshoeff:1997kr}.
Consider the two Majorana--Weyl supersymmetry parameters of type IIA, $\epsilon_1$ and $\epsilon_2$. An $O_p$-plane/$D_p$-brane source introduces a relation between them of the type
\begin{equation}
\epsilon_1=\Gamma_{(p)}\epsilon_2\,,
\label{eq:e1Ge2}
\end{equation}
where $\Gamma_{(p)}$ is a product of ten-dimensional gamma matrices with legs along the source worldvolume, possibly with appropriate orientation sign and chirality factors understood; we assume vanishing worldvolume flux in what follows. Such a relation halves the original amount of supersymmetry and we refer to it as a projection.
An additional $O_p/D_p$ source introduces an analogous projection with $\Gamma_{(p)}'$.
For orthogonal sources, two independent source projections preserve one quarter of the original supersymmetry if $[\Gamma_{(p)},\Gamma_{(p)}']=0$, and each additional independent projection commuting with all previous ones halves the remaining supersymmetry, always provided the orientation signs are mutually compatible.

In the setup of \cite{Cribiori:2026caf}, all sources fill $\mathrm{AdS}_3$ and wrap a four-cycle in the seven internal directions $1,\dots,7$:
\begin{equation}
 \mathrm{O6}:\quad 1237,\;1567,\;2467,\;3457,
 \qquad
 \mathrm{D6}:\quad 1245,\;1346,\;2356.
 \label{eq:planes}
\end{equation}
We label these four-cycles by $I,J,\dots$ and use gamma matrices in an orthonormal frame, so that, for example, $\Gamma_{1237}=\Gamma_1\Gamma_2\Gamma_3\Gamma_7$. The condition \eqref{eq:e1Ge2} on the ten-dimensional spinors becomes
\begin{equation}
\epsilon_1=\sigma_I\Gamma_{\mathrm{AdS}_3}\Gamma_I\,\epsilon_2,
\end{equation}
where $\sigma_I=\pm1$ encodes the source orientation and $\Gamma_{\mathrm{AdS}_3}$ is the product of the ten-dimensional gamma matrices in the external directions.
For all sources in \eqref{eq:planes}, one can check that $[\Gamma_I,\Gamma_J]=0$, so the full source projections also commute. Assuming compatible orientation signs (if the signs are not compatible it means of course directly that spontaneous breaking takes place), it remains to count the independent source projections.  

To count these projections, we first use the O6 condition along $1237$ to express $\epsilon_1$ in terms of $\epsilon_2$. Each remaining source condition then becomes
\begin{equation}
\Gamma_{1237}\Gamma_I\epsilon_2
=\sigma_{1237}\sigma_I\epsilon_2.
\end{equation}
In the O6-plane sector, the conditions along $1567$ and $2467$ give two independent projections on $\epsilon_2$, since $\Gamma_{1237}\Gamma_{1567}=-\Gamma_{2356}$ and $\Gamma_{1237}\Gamma_{2467}=\Gamma_{1346}$. The condition along $3457$ is not independent but follows from these two for appropriate orientation signs, because $\Gamma_{1237}\Gamma_{3457} =\Gamma_{1567}\Gamma_{2467} =-\Gamma_{1245}$, while $\Gamma_{2356}\Gamma_{1346}=\Gamma_{1245}$. Including the first condition relating $\epsilon_1$ to $\epsilon_2$, there are therefore three independent source projections. The four O6-planes thus leave $32/2^3=4$ supercharges, consistent with the $\mathcal{N}=(1,1)$ supersymmetry of the bulk setup.

The D6-branes introduce one additional independent projection. Indeed, the condition along $1245$ involves $\Gamma_{3457}$, since $\Gamma_{1237}\Gamma_{1245}=-\Gamma_{3457}$. This matrix commutes with $\Gamma_{2356}$ and $\Gamma_{1346}$ but cannot be generated by their products, since neither contains $\Gamma_7$. It therefore imposes an independent projection on $\epsilon_2$.
For compatible orientation signs, the other two D6 conditions follow from products of the independent ones. The D6-branes therefore halve the remaining supersymmetry only once, leaving $4/2=2$ supercharges, consistent with $\mathcal{N}=1$.
Since D-branes do not impose an orientifold truncation of the gravitini, we expect this breaking to be spontaneous, with the broken supersymmetry non-linearly realized. Establishing this interpretation requires including explicitly the open string sector in the three-dimensional effective theory.

We stress that this counting applies to the source conditions in the toroidal orbifold limit with smeared backreaction. On a smooth resolution with fully localized sources, the corresponding projectors would have to be determined anew, but we are unable to perform this analysis. We therefore leave open the amount of preserved supersymmetry in that setting, including the possibility of having $\mathcal{N}=(1,1)$ fully linearly realized.

\section{Ten-dimensional solution}

We seek supersymmetric solutions with three-dimensional anti-de Sitter external space and internal space $\mathcal{M}_7$. We first approach the problem from the ten-dimensional perspective by imposing the vanishing of the type IIA fermionic supersymmetry variations. The simplest spinor ansatz compatible with minimal supersymmetry in three dimensions takes the two internal supersymmetry spinors to be everywhere parallel, so that they define a strict $G_2$ structure. We first prove that supersymmetric, three-dimensional anti-de Sitter solutions of this type cannot exist in massless type IIA supergravity. Motivated by this obstruction, we then consider non-parallel internal spinors, whose common stabilizer is $\mathrm{SU}(3)\subset G_2$.

\subsection{No $G_2$ anti-de Sitter solutions in massless IIA supergravity}

We first prove that three-dimensional supersymmetric anti-de Sitter solutions with a strict $G_2$ structure cannot exist in massless type IIA supergravity. Our strategy is to identify an obstruction in the ten-dimensional Killing-spinor equations. From a purely technical perspective, it is convenient to express these equations in terms of differential forms constructed from spinor bilinears, usually referred to as bispinor equations. Their form can be found $e.g.$ in \cite{Macpherson:2021lbr}, which we follow up to some minor differences in conventions that we will specify in due time. 

We consider the string-frame metric
\begin{equation}
    ds^2_{10}=e^{2A}ds^2_{\mathrm{AdS}_3}
    +ds^2_{\mathcal M_7}\,,
\end{equation}
where the warp factor $A$, and likewise the dilaton $\phi$, depend only on the internal coordinates; the unwarped anti-de Sitter metric has radius $L_{\rm AdS}=|\mu|^{-1}$.
For minimal supersymmetry in three dimensions, the two ten-dimensional type IIA Majorana--Weyl supersymmetry parameters decompose as
\begin{equation}
    \epsilon_1=\zeta\otimes\theta_+\otimes\chi_1\,,
    \qquad
    \epsilon_2=\zeta\otimes\theta_-\otimes\chi_2\,.
    \label{eq:eps12IIA}
\end{equation}
Here, $\zeta$ is a common three-dimensional Majorana Killing spinor, $\chi_1,\chi_2$ are internal seven-dimensional Majorana spinors, and $\theta_\pm$ are fixed auxiliary two-component spinors encoding the opposite ten-dimensional chiralities. In type IIA, and in the basis used in \cite{Macpherson:2021lbr}, they take the form $\theta_\pm=\frac{1}{\sqrt2}\left(\begin{smallmatrix}1\\ \mp i\end{smallmatrix}\right)$.
Supersymmetry further imposes the relations
\begin{equation}
    \|\chi_1\|^2+\|\chi_2\|^2=c_+e^A\,,
    \qquad
    \|\chi_1\|^2-\|\chi_2\|^2=c_-e^{-A}\,,
    \label{eq:spinor-norms}
\end{equation}
where $c_+>0$ and $c_-$ are constants, and, for later use, it is convenient to introduce the quantities
\begin{equation}
    \rho=\sqrt{1-\frac{c_-^2}{c_+^2}e^{-4A}}\,,
    \qquad
    n=\|\chi_1\|\,\|\chi_2\|
    =\frac{c_+}{2}e^A\rho\,.
\end{equation}
Assuming both internal spinors are nonzero, we have $0<\rho\leq1$, with $\rho=1$, equivalently $c_-=0$, corresponding to equal norms.

We take $\zeta$ to satisfy the three-dimensional Killing-spinor equation, so we concentrate on the internal spinors $\chi_1$ and $\chi_2$.
Recall that a $G_2$ structure on a seven-dimensional Riemannian spin manifold is specified by a Majorana spinor whose bilinear defines a nowhere-vanishing three-form $\Phi$ such that $\Phi\wedge *_7\Phi=7\,\mathrm{vol}_7$, with ${\rm vol}_7$ the seven-dimensional volume form. 
A supersymmetric solution of type II supergravity compactified on such seven-dimensional manifold is said to be of (strict) $G_2$ type if $\chi_1$ and $\chi_2$ define the same $G_2$ structure and are therefore parallel, although their norms may differ. Then, given the bispinor $\chi_1\otimes\chi_2^\dagger$, the Clifford map identifies the terms in its Fierz expansion with differential forms. We group its real even- and odd-degree representatives into two polyforms, denoted in our conventions by $\Psi_+^{(G_2)}$ and $\Psi_-^{(G_2)}$, respectively.
For the strict $G_2$ ansatz, in our conventions they read
\begin{equation}
\label{eq:G2bispinors}
    \Psi_+^{(G_2)}=\frac n8\left(1-*_7\Phi\right),
    \qquad
    \Psi_-^{(G_2)}=\frac n8\left(\Phi-\mathrm{vol}_7\right),
\end{equation}
and satisfy
\begin{equation}
    *_7\lambda(\Psi_\pm^{(G_2)})=\mp\Psi_\mp^{(G_2)},
    \qquad
    \langle\Psi_+^{(G_2)},\Psi_-^{(G_2)}\rangle
    =\frac{n^2}{8}\,\mathrm{vol}_7,
\end{equation}
where $\lambda(A_p)=(-1)^{\lfloor \frac p2 \rfloor}A_p$, and the seven-dimensional Mukai pairing is $\langle \alpha ,\beta\rangle=[\alpha\wedge\lambda(\beta)]_7$, with $\alpha$, $\beta$ polyforms and with $[\cdot]_7$ extracting the seven-form component.

To keep the argument general, we allow both internal and external NS flux,
\begin{equation}
    H_{10}=H_3+e^{3A}h_0\,\mathrm{vol}_{\mathrm{AdS}_3}\,,
\end{equation}
where the external volume form refers to the unwarped metric.
The supersymmetry conditions yielding \eqref{eq:spinor-norms} also give the relation
\begin{equation}
    c_+e^{3A}h_0=2\mu c_-\,.
    \label{eq:electric-norm}
\end{equation}
We denote the internal RR polyform by $F=F_0+F_2+F_4+F_6$ and define $d_H=d-H_3\wedge$.
The necessary bispinor equations given in \cite{Macpherson:2021lbr}, once translated to our conventions, read\footnote{With respect to \cite{Macpherson:2021lbr} we send $\psi_- \to - \Psi_-$, $m \to - \mu$, and $f \to -F$ in type IIB.}
\begin{align}
\label{eq:bispineq1}
    d_H\left(e^{A-\phi}\Psi_\mp\right)
    &=\pm\frac{1}{16}c_-F\,,
    \\
    \label{eq:bispineq2}
    d_H\left(e^{2A-\phi}\Psi_\pm\right)
    \mp 2\mu e^{A-\phi}\Psi_\mp
    &=\frac{1}{16}c_+e^{3A}*_7\lambda(F)\,,
    \\
    \label{eq:bispineq3}
    \langle\Psi_\mp,F\rangle
    &=\mp\frac14 c_+\mu\rho^2e^{-\phi}\mathrm{vol}_7\,,
\end{align}
where the upper/lower sign is for type IIA/B respectively. Their solution for $\Psi_\pm = \Psi_\pm^{(G_2)}$ is 
\begin{subequations}
\begin{align}
c_- F_0&=0,\\
c_- F_2 &=0,\\
d(e^{2A-\phi}c_+\rho\, \Phi) - c_- F_4 &=0,\\
c_+e^{2A-\phi}\rho \,H_3 \wedge \Phi + c_-F_6&=0,\\[.2cm]
e^{3A}*_7 F_6 &= - d(e^{3A-\phi} \rho),\\
e^{3A}*_7 F_4 &=-e^{3A-\phi}\rho H_3-2 \mu \rho\, e^{2A-\phi}\Phi,
\label{eq:G2bispinorF4}\\
e^{3A}*_7 F_2 &=  d(e^{3A-\phi}\rho *_7 \Phi),\\
e^{3A}*_7 F_0 &= e^{3A-\phi} \rho \, H_3 \wedge *_7 \Phi + 2\mu \rho e^{2A-\phi}{\rm vol}_7,
\label{eq:G2bispinorF0}\\[.2cm]
\Phi \wedge F_4 - {\rm vol}_7 F_0 &= -4 \mu \rho \,e^{-A-\phi}{\rm vol}_7,
\label{eq:G2bispinorpairIIA}
\end{align}
\end{subequations}
for type IIA, and
\begin{subequations}
\begin{align}
d(e^{2A-\phi}c_+\rho) + c_- F_1 &=0,\\
e^{2A-\phi}c_+\rho H_3 - c_- F_3 &=0,\\
d(e^{2A-\phi}\rho c_+ *_7 \Phi) -c_- F_5 &=0,\\
e^{2A-\phi} \rho c_+ H_3 \wedge*_7 \Phi -c_- F_7 &=0,\\[.2cm]
e^{3A}*_7 F_7 &=- 2 \mu \rho\, e^{2A-\phi},\\
e^{3A} *_7 F_5 &=0,\\
e^{3A} *_7 F_3 &=- d(e^{3A-\phi} \rho \Phi) + 2 \mu \rho\, e^{2A-\phi}  *_7 \Phi,\\
e^{3A}*_7F_1 &= - e^{3A-\phi}\rho H_3 \wedge \Phi,\\[.2cm]
*_7 \Phi \wedge F_3 - F_7 &= 4 \mu \rho\, e^{-A-\phi}{\rm vol}_7,
\end{align}
\end{subequations}
for type IIB; the latter is reported here only for completeness. 

We are now ready to find the sought after obstruction. We first act on \eqref{eq:G2bispinorF4} with $\Phi \wedge *_7$ to get 
\begin{equation}
-14 \mu \rho\, {\rm vol}_7 = e^{A+\phi}\Phi \wedge F_4 + \rho e^A *_7 \Phi \wedge H_3.
\end{equation}
Then, we subtract from this both 
\eqref{eq:G2bispinorF0} and \eqref{eq:G2bispinorpairIIA} to find
\begin{equation}
e^{A+\phi}{\rm vol}_7 F_0 = -4\mu \rho\, {\rm vol}_7.
\end{equation}
Together with \eqref{eq:G2bispinorF0}, this implies the relations
\begin{equation}
e^{A+\phi}F_0 = - 4 \mu \rho, \qquad H_3 \wedge *_7 \Phi = -6 \mu e^{-A} {\rm vol}_7.
\end{equation}
Since by construction $\rho > 0$, we see that the first of these equations forces $\mu=0$ whenever $F_0=0$. Hence, we conclude that there cannot be supersymmetric anti-de Sitter solutions of (strict) $G_2$ type in massless type IIA. Notice that this conclusion was already reached in \cite{VanHemelryck:2022ynr} from the same bispinor equations but in the subcase $c_-=0$, implying that the spinors $\chi_1$, $\chi_2$ have equal norm. Here, we extended it to arbitrary $c_-$, namely to the case in which the norms of $\chi_1$, $\chi_2$ are not necessarily the same.

\subsection{SU(3) structure solutions}
\label{sec:SU3solution}

Motivated by the result of the previous section, we try to find solutions with SU$(3)\subset G_2$ structure. They can still give rise to minimal supersymmetry in three dimensions now realized as in \eqref{eq:eps12IIA} but with $\chi_1$, $\chi_2$ not parallel but orthogonal to each other. (The more general case in which the angle between $\chi_1$ and $\chi_2$ is not 0 nor $\pi/2$ is usually called dynamic SU$(3)$ or $G_2 \times G_2$ structure, see $e.g.$~\cite{Passias:2020ubv}.)

A $G_2$ structure decomposes the eight-dimensional spinor representation of $\mathrm{Spin}(7)$ as $\mathbf8\to\mathbf1\oplus\mathbf7$. Hence, invariant internal spinors must be parallel, since there is only one singlet. An $\mathrm{SU}(3)$ structure further decomposes $\mathbf7\to\mathbf1\oplus\mathbf3\oplus\bar{\mathbf3}$, so that the spinor representation now contains two singlets, $\mathbf8\to\mathbf1\oplus\mathbf1\oplus\mathbf3 \oplus\bar{\mathbf3}$ (the decompositions involving $\mathbf3$ and $\bar{\mathbf3}$ are understood after complexification; together, they describe a six-dimensional real representation). The two invariant real spinors can therefore be chosen orthogonal.

Passing from a $G_2$ to an $\mathrm{SU}(3)$ structure amounts to choosing a distinguished unit vector, since the subgroup of $G_2$ preserving such a vector is $\mathrm{SU}(3)$ \cite{chiossi2002}. We denote its dual one-form by $v$, consistently with the notation of the previous sections. The six-dimensional subspace orthogonal to this vector is naturally described by the $\mathrm{SU}(3)$-structure forms $J$ and $\Omega$ previously introduced. In our setup, $v=e^4$, and this subspace is tangent to the six-dimensional factor of $\mathcal M_7$. The SU$(3)$-structure forms are related to $\Phi$ and $v$ by \cite{chiossi2002}
\begin{equation}
    J=\iota_{v^\sharp}\Phi,
    \qquad
    \mathrm{Re}\Omega=\Phi-v\wedge \iota_{v^\sharp}\Phi,
    \qquad
    \mathrm{Im}\Omega = -\iota_{v^\sharp}(*_7\Phi),
\end{equation}
where $v^\sharp$ is the vector dual to $v$, or equivalently
\begin{equation}
\Phi = v \wedge J + \mathrm{Re}\Omega, \qquad *_7\Phi=\frac12 J\wedge J-v\wedge\mathrm{Im}\Omega.
\end{equation}
In these expressions we are using the orientation $\mathrm{vol}_7=v\wedge\mathrm{vol}_6$, with ${\rm vol_6}=\frac{1}{6}J \wedge J \wedge J$.

Simply substituting these expressions into \eqref{eq:G2bispinors} would only rewrite the strict $G_2$ bispinors in $\mathrm{SU}(3)$ language, without departing from the case in which $\chi_1$ and $\chi_2$ are parallel. To construct orthogonal spinors, we start from the unit spinor $\chi$ defining $\Phi$ and set $\chi_1=-i\,\|\chi_1\|\, v_m\gamma^m\chi$ and $\chi_2=\|\chi_2\|\chi$, using imaginary antisymmetric gamma matrices satisfying $\{\gamma^m,\gamma^n\}=2g^{mn}$. These spinors are orthogonal since $\chi^T\gamma^m\chi=0$, and their norms are unchanged from the previous case because $v$ is unit; the factor of $i$ ensures that $\chi_1$ remains real. Then, the Clifford map applied to the Fierz expansion of $\chi_1\otimes\chi_2^\dagger$ gives the desired polyforms $\Psi_\pm^{(\mathrm{SU}(3))}$. Indeed, since the mixed bispinor is obtained by left Clifford multiplication of $n\,\chi\otimes\chi^\dagger$ by $-i\slashed v$, its form representatives follow by applying $\pm(v\wedge+\iota_{v^\sharp})$ to the $G_2$ polyforms \eqref{eq:G2bispinors}.\footnote{Under the Clifford map, left multiplication by $\slashed v$ corresponds to $v\wedge+\iota_{v^\sharp}$. This follows from $\gamma^m\gamma^{n_1\ldots n_p} =\gamma^{mn_1\ldots n_p} +p\,g^{m[n_1}\gamma^{n_2\ldots n_p]}$; in particular, $\slashed v\slashed u =v_m u_n\gamma^{mn}+v\cdot u$.} Since the action of $\pm(v\wedge+\iota_{v^\sharp})$ exchanges even and odd degree, we obtain \cite{Passias:2020ubv}
\begin{equation}
    \Psi_+^{(\mathrm{SU}(3))}
    =-\left(v\wedge+\iota_{v^\sharp}\right)
      \Psi_-^{(G_2)},
    \qquad
    \Psi_-^{(\mathrm{SU}(3))}
    =\left(v\wedge+\iota_{v^\sharp}\right)
      \Psi_+^{(G_2)}.
\end{equation}
Explicitly, these read
\begin{equation}
    \Psi_+^{(\mathrm{SU}(3))}
    =-\frac n8\left(
        J+v\wedge\mathrm{Re}\Omega-\mathrm{vol}_6
    \right),
    \qquad
    \Psi_-^{(\mathrm{SU}(3))}
    =\frac n8\left(
        v+\mathrm{Im}\Omega-\frac12v\wedge J\wedge J
    \right).
\end{equation}
The relative signs can be fixed by $*_7\lambda(\Psi_\pm)=\mp\Psi_\mp$, while the common overall sign is fixed by our choice $\chi_1=-i\,\|\chi_1\|\,\slashed v\,\chi$ and $\chi_2=\|\chi_2\|\chi$; besides these polyforms obey also $\langle\Psi_+,\Psi_-\rangle=\frac{n^2}{8}\mathrm{vol}_7$.

We now substitute these expressions into the bispinor equations \eqref{eq:bispineq1}-\eqref{eq:bispineq3} for type IIA. We take the warp factor and dilaton to be constant, absorbing the former into the external metric so that $A=0$, and denote the string coupling by $g_s=e^\phi$. We also impose $H_{10}=0$ and $F_0=0$, as appropriate to our setup. Since $\mu\neq0$, the algebraic condition \eqref{eq:electric-norm} implies $c_-=0$, and hence $\rho=1$ and $n=c_+/2$. The common spinor normalization remains arbitrary; for convenience, we choose $c_+=2$, so that $n=1$. 
The type IIA supersymmetry equations then reduce to
\begin{subequations}
\begin{align}
    d\Psi_-&=0,\\
    d\Psi_+-2\mu\Psi_-
    &=\frac{g_s}{8}*_7\lambda(F),\\
    \langle\Psi_-,F\rangle
    &=-\frac{\mu}{2g_s}\mathrm{vol}_7.
    \label{eq:pairingeqSU3IIA}
\end{align}
\end{subequations}

For the geometry described in section \ref{sec:geometry}, $dv=0$, $d\,\mathrm{Im}\Omega=0$ and $d(J\wedge J)=0$, so the first equation is automatically satisfied. Separating the second equation by form degree gives
\begin{subequations}
\begin{align}
    g_s *_7F_6&=2\mu v,\\
    g_s *_7F_4&=-dJ-2\mu\,\mathrm{Im}\Omega,\\
    g_s *_7F_2&=-v\wedge
        \left(d\,\mathrm{Re}\Omega+\mu J\wedge J\right),
\end{align}
\end{subequations}
and using the torsion classes defined in \eqref{eq:torsion-classes} leads to
\begin{subequations}
\begin{align}
g_s F_2 &= -2(\mu + w_1)J+w_2\, ,\\
g_s F_4 &=-\left(\frac32 w_1+2\mu\right)v \wedge \mathrm{Re}\Omega + v \wedge *_6 w_3\, ,\\
g_s F_6 &= 2 \mu {\rm vol}_6.
\end{align}
\label{eq:SU3-RR-fluxes}
\end{subequations}
The remaining pairing equation \eqref{eq:pairingeqSU3IIA} reduces then to
\begin{equation}
\mu = - w_1,
\end{equation}
which allows to further simplify the RR fluxes as
\begin{subequations}
\begin{align}
g_s F_2 &= w_2\, ,\\
g_s F_4 &=\frac12 w_1\,v \wedge \mathrm{Re}\Omega + v \wedge *_6 w_3\, ,\\
g_s F_6 &= -2w_1 {\rm vol}_6.
\end{align}
\label{eq:SU3-RR-fluxes2}
\end{subequations}
Recalling that $w_1$, $w_2$ and $w_3$ in our model are given explicitly in \eqref{eq:w1}, \eqref{eq:w2} and \eqref{eq:w3} and matching the above expressions to the flux expansions \eqref{RR2_expansion}-\eqref{RR6_expansion} we identify
\begin{subequations}
\begin{align}
    m_0&=\frac{2\Delta}{3g_s}r_1r_5,\\
    N&=-\frac{\Delta}{3g_s}r_2r_6,\\
    R_F&=-\frac{\Delta}{3g_s}r_3r_7,\\[.2cm]
    f_1&=-\frac{a_1-\Delta/3}{g_s}
        r_4r_1r_3r_6,\\
    f_2&=\frac{a_3+\Delta/3}{g_s}
        r_4r_2r_3r_5,\\
    f_3&=-\frac{a_2-\Delta/3}{g_s}
        r_4r_1r_2r_7,\\
    f_4&=\frac{a_4+\Delta/3}{g_s}
        r_4r_5r_6r_7,\\[.2cm]
            M&=-\frac{\Delta}{3g_s}
        r_1r_2r_3r_5r_6r_7,\\[.2cm]
L_{\rm AdS} &= 1/|\mu|=6 /|\Delta|.
\end{align}
\label{eq:flux-radius-relations}
\end{subequations}

We now express the solution in terms of the scalars $t_i$ and $u_i$
defined in \eqref{eq:scalarsmodel}. Taking ratios of the two-form
and six-form flux coefficients gives
\begin{equation}
    t_1=\sqrt{-\frac{m_0M}{2NR_F}},
    \qquad
    t_2=\sqrt{-\frac{2NM}{m_0R_F}},
    \qquad
    t_3=\sqrt{-\frac{2R_FM}{m_0N}}.
    \label{eq:t-flux-solution}
\end{equation}
All square roots are taken to be positive. Positivity of the radii
and of $g_s$ requires $m_0$ to have the same sign as $h$, while
$N$, $R_F$ and $M$ have the opposite sign.
To determine the shape moduli, we introduce
\begin{equation}
(1+z_1,1+z_2,1+z_3,1+z_4) =\frac{6}{\Delta}(a_1,-a_3,a_2,-a_4),  \qquad   \sum_{i=1}^4z_i=2.
\label{eq:zidef}
\end{equation}
The four-form flux parameters then can be recast as
\begin{equation}
f_i=C(1-z_i^2), \qquad C=\frac{r_4t_2t_3\Delta^2}{36hg_s},
    \label{eq:f-z}
\end{equation}
which can be inverted as
\begin{equation}
    z_i=\varepsilon_i\sqrt{1-\frac{f_i}{C}},
    \qquad
    \varepsilon_i=\pm1\,.
    \label{eq:z-flux}
\end{equation}
Substituting these expressions into $\sum_i z_i=2$ determines
the combination of moduli $C$ in terms of the fluxes through
the single algebraic equation
\begin{equation}
    \sum_{i=1}^4
        \varepsilon_i\sqrt{1-\frac{f_i}{C}}=2.
    \label{eq:C-flux}
\end{equation}
For each choice of signs, we retain only finite nonzero real roots for which all square roots are real, $z_i>-1$ and $C/h>0$. The fluxes must additionally satisfy the $F_4$ Bianchi identity \eqref{eq:dF4tad}, namely $\sum_i f_i=0$, which, using \eqref{eq:f-z} and $C\neq0$, is equivalent to 
\begin{equation}
\sum_i z_i^2=4.
\end{equation}
We have already solved the $F_2$ Bianchi identity in section~\ref{sec:fluxes}, while the $F_6$ Bianchi identity is automatically satisfied in our ansatz.
The $u_i$ scalars follow directly from the definitions of $a_i$ in \eqref{ai_expr},
\begin{equation}
\begin{aligned}
    u_1&=\sqrt{\frac{a_3a_4}{a_1a_2}}
        =\sqrt{\frac{(1+z_2)(1+z_4)}
                         {(1+z_1)(1+z_3)}},\\
    u_2&=\sqrt{\frac{a_1a_4}{a_2a_3}}
        =\sqrt{\frac{(1+z_1)(1+z_4)}
                         {(1+z_3)(1+z_2)}},\\
    u_3&=\sqrt{\frac{a_2a_4}{a_1a_3}}
        =\sqrt{\frac{(1+z_3)(1+z_4)}
                         {(1+z_1)(1+z_2)}}.
\end{aligned}
    \label{eq:u-flux-solution}
\end{equation}
Equivalently, the six radii entering $t_i$ and $u_i$ are
\begin{equation}
 r_1=\sqrt{\frac{t_1}{u_1}}, \quad r_2=\sqrt{\frac{t_2}{u_2}}, \quad  r_3=\sqrt{\frac{t_3}{u_3}}, \quad  r_5=\sqrt{t_1u_1}, \quad r_6=\sqrt{t_2u_2}, \quad r_7=\sqrt{t_3u_3}.
 \label{eq:6radiitu}
\end{equation}
To express the remaining quantities in terms of fluxes, it is convenient to introduce $\mathcal Z=\prod_{i=1}^4(1+z_i)$. Then, using $\Delta=6h\,\mathcal Z^{-1/4}\sqrt{t_1/(t_2t_3)}$, we obtain
\begin{equation}
\begin{aligned}
    r_4&=\frac{4C\mathcal Z^{1/4}}{m_0}
        \sqrt{\frac{t_1}{t_2t_3}},\\
    g_s&=\frac{4h}{m_0\mathcal Z^{1/4}}
        \sqrt{\frac{t_1^3}{t_2t_3}},\\
    L_{\rm AdS}&=\frac{\mathcal Z^{1/4}}{|h|}
        \sqrt{\frac{t_2t_3}{t_1}}.
\end{aligned}
    \label{eq:remaining-flux-solution}
\end{equation}
For each admissible branch of \eqref{eq:C-flux}, all metric moduli and the dilaton are expressed in terms of fluxes and thus stabilized; the quantities $t_i$, $z_i$, $\mathcal Z$ and $C$ appearing above are shorthand for the flux-dependent expressions just determined.

We can obtain a particularly simple solution by choosing $h>0$ and then restricting to
\begin{equation}
    f_1=f_2=-f,
    \qquad
    f_3=f_4=f,
    \qquad
    f>0.
    \label{eq:symmsolf}
\end{equation}
The $F_4$ Bianchi identity is thus trivially satisfied and we find\footnote{Since $h>0$ we have $C>0$. The $F_4$ flux equations give $z_1^2=z_2^2= 1+ \frac fC$, $z_3^2 = z_4^2= 1-\frac fC$. Since $\sqrt{1+\frac fC}>1$, the condition $z_i>-1$ forces $z_1 = z_2 = \sqrt{1+\frac fC}$ and $z_3 = z_4 = -\sqrt{1-\frac fC}$. The algebraic equation \eqref{eq:C-flux} becomes then $\sqrt{1+\frac fC}-\sqrt{1-\frac fC}=1$, whose positive solution is $C= 2f/\sqrt 3$.}
\begin{equation}
    C=\frac{2f}{\sqrt3},
    \qquad
    z_1=z_2=\frac{1+\sqrt3}{2},
    \qquad
    z_3=z_4=\frac{1-\sqrt3}{2},
\end{equation}
giving
\begin{equation}
    \mathcal Z=\frac94,
    \qquad
    u_1=u_2=1,
    \qquad
    u_3=2-\sqrt3 , 
\end{equation}
and also $r_1=r_5$, $r_2=r_6$ and $r_3/r_7=2+\sqrt3$.
To display the solution directly in terms of positive flux parameters, we introduce $\widehat N=-N>0$, $\widehat R=-R_F>0$ and $\widehat M=-M>0$, with $m_0>0$.  The area moduli are
\begin{equation}
    t_1=\sqrt{\frac{m_0\widehat M}
                    {2\widehat N\widehat R}},
    \qquad
    t_2=\sqrt{\frac{2\widehat N\widehat M}
                    {m_0\widehat R}},
    \qquad
    t_3=\sqrt{\frac{2\widehat R\widehat M}
                    {m_0\widehat N}},
\end{equation}
and all seven radii read
\begin{equation}
\begin{aligned}
    r_1=r_5
    &=\left(\frac{m_0\widehat M}
                   {2\widehat N\widehat R}\right)^{1/4},\\
    r_2=r_6
    &=\left(\frac{2\widehat N\widehat M}
                   {m_0\widehat R}\right)^{1/4},\\
    r_3
    &=\sqrt{2+\sqrt3}\,
      \left(\frac{2\widehat R\widehat M}
                   {m_0\widehat N}\right)^{1/4},\\
    r_7
    &=\sqrt{2-\sqrt3}\,
      \left(\frac{2\widehat R\widehat M}
                   {m_0\widehat N}\right)^{1/4},\\
    r_4
    &=\frac{2^{7/4}f}
      {(m_0\widehat M\widehat N\widehat R)^{1/4}}.
\end{aligned}
\end{equation}
The string coupling and the AdS radius are
\begin{equation}
    g_s=\frac{2^{5/4}}{\sqrt3}\,h
       \left(\frac{m_0\widehat M}
       {\widehat N^3\widehat R^3}\right)^{1/4},
    \qquad
    L_{\rm AdS}=\frac{2^{1/4}\sqrt3}{h}
       \left(\frac{\widehat M\widehat N\widehat R}
       {m_0^3}\right)^{1/4}.
\end{equation}

\subsection{Parametric scale separation}
\label{sec:scale-separation}

We now investigate whether the solutions above admit parametric scale separation at large internal volume and weak or strong string coupling. We keep the metric flux $h$ and the two-form flux parameter $m_0$ fixed, consistently with the tadpole condition \eqref{eq:F2tadheven}, and scale the remaining RR fluxes. For definiteness, we take $h,m_0>0$ and use the positive parameters $\widehat N=-N$, $\widehat R=-R_F$ and $\widehat M=-M$.

The following analysis applies to any fixed admissible branch of the algebraic equation \eqref{eq:C-flux}. Indeed, under a common rescaling $f_i=\lambda^\delta\bar f_i$, with $\delta \in \mathbb{R}$ and $\sum_i\bar f_i=0$, the equations
$f_i=C(1-z_i^2)$ remain satisfied by $C=\lambda^\delta\bar C$ and fixed $z_i=\bar z_i$; consequently, the shape moduli $u_i$ in \eqref{eq:u-flux-solution} and the quantity $\mathcal Z$ remain fixed.\footnote{A more general scaling analysis can involve different exponents $\delta_i$. In such a case, the scalars $u_i$ would not remain fixed in general.} 
We assume that this branch is nondegenerate, so that all $1+\bar z_i$ are strictly positive. The symmetric solution
$f_1=f_2=-f$, $f_3=f_4=f$ provides an explicit example, but is not
required for the scaling argument, which is more general.

Introduce $\alpha, \beta, \gamma, \delta \in \mathbb{R}$, a scaling parameter $\lambda$ and let
\begin{equation}
    \widehat N\sim\lambda^\alpha,
    \qquad
    \widehat R\sim\lambda^\beta,
    \qquad
    \widehat M\sim\lambda^\gamma,
    \qquad
    f_i=\lambda^\delta \bar f_i,
    \qquad
    \lambda\to\infty,
    \label{eq:flux-scaling}
\end{equation}
with the limit being understood along sequences compatible with flux quantization; here $\sim$ denotes equality up to nonzero factors independent of $\lambda$. The expressions \eqref{eq:t-flux-solution} give
\begin{equation}
    t_1\sim\lambda^{\frac12(\gamma-\alpha-\beta)},
    \qquad
    t_2\sim\lambda^{\frac12(\gamma+\alpha-\beta)},
    \qquad
    t_3\sim\lambda^{\frac12(\gamma-\alpha+\beta)}.
    \label{eq:area-moduli-scaling}
\end{equation} 
Since the $u_i=\bar u_i>0$ remain fixed, the six radii of the twisted torus in \eqref{eq:6radiitu} scale as 
\begin{equation}
r_1 \sim r_5 \sim \lambda^{\frac14(\gamma-\alpha-\beta)}, \qquad r_2 \sim r_6 \sim \lambda^{\frac14(\gamma+\alpha-\beta)}, \qquad r_3 \sim r_7 \sim \lambda^{\frac14(\gamma-\alpha+\beta)}. 
\label{eq:radii-scaling}
\end{equation}
Notice that, while their ratios are fixed by $\bar u_i$, and their scaling exponents coincide, the above paired radii need not be equal.
Using $C=\lambda^\delta\bar C$ and $\mathcal Z=\bar{\mathcal Z}$ in \eqref{eq:remaining-flux-solution}, we similarly obtain 
\begin{equation}
\begin{aligned}
r_4&  \sim\lambda^{\delta-\frac14(\gamma+\alpha+\beta)},\\
 g_s&\sim\lambda^{\frac14(\gamma-3\alpha-3\beta)},\\
L_{\rm AdS}&  \sim\lambda^{\frac14(\gamma+\alpha+\beta)}.
\end{aligned}
    \label{eq:coupling-AdS-scaling}
\end{equation}
The seven-dimensional volume in string units therefore scales as
\begin{equation}
    {\rm vol}_7\sim r_4t_1t_2t_3
    \sim\lambda^{\delta+\frac14(5\gamma-3\alpha-3\beta)}.
\end{equation}
These scalings apply to any nondegenerate solution of \eqref{eq:C-flux}, independently of whether the four-form fluxes obey the symmetric choice \eqref{eq:symmsolf}.
We will impose that every internal radius becomes large, which is stronger than requiring only a large total volume. This gives
\begin{equation}
\label{eq:largevolumes}
    \gamma>\max\{\alpha+\beta,|\alpha-\beta|\},
    \qquad
    \delta>\frac{\gamma+\alpha+\beta}{4}.
\end{equation}

To investigate scale separation, we compare each internal radius with $L_{\rm AdS}$. In the approximation discussed in section~\ref{sec:fluxes}, the geometric Kaluza-Klein scales are set by inverse radii, up to factors that do not affect the parametric analysis. The relevant dimensionless ratios scale as
\begin{equation}
\begin{aligned}
    \frac{r_1^2}{L_{\rm AdS}^2}\sim \frac{r_5^2}{L_{\rm AdS}^2}
    &\sim\lambda^{-\alpha-\beta}\,,\qquad \frac{r_2^2}{L_{\rm AdS}^2}\sim\frac{r_6^2}{L_{\rm AdS}^2}
    \sim\lambda^{-\beta},\\
    \frac{r_3^2}{L_{\rm AdS}^2}\sim\frac{r_7^2}{L_{\rm AdS}^2}
    &\sim\lambda^{-\alpha}\,,\qquad \qquad\qquad\,\,\,\,\,\frac{r_4^2}{L_{\rm AdS}^2}
    \sim\lambda^{2\delta-\gamma-\alpha-\beta},
\end{aligned}
    \label{eq:scale-separation-ratios}
\end{equation}
thus all seven radii become parametrically smaller than
$L_{\rm AdS}$ if
\begin{equation}
    \alpha>0,
    \qquad
    \beta>0,
    \qquad
    \delta<\frac{\gamma+\alpha+\beta}{2}.
\end{equation}
Combining these with the conditions \eqref{eq:largevolumes} giving large internal radii results in
\begin{equation}
    \alpha,\beta>0,
    \qquad
    \gamma>\alpha+\beta,
    \qquad
    \frac{\gamma+\alpha+\beta}{4}
    <\delta<
    \frac{\gamma+\alpha+\beta}{2}.
    \label{eq:large-radius-scale-separation}
\end{equation}
We have thus shown that our solutions admit parametric scale separation with all internal radii parametrically large in string units. Since $z_i=\bar z_i$ remain fixed, their definition in \eqref{eq:zidef} gives $|a_i|\sim|\Delta|\sim L_{\rm AdS}^{-1}$.\footnote{The components of the Riemann tensor of the left-invariant metric in the orthonormal frame, being quadratic in the $a_i$, therefore also vanish in string units in this limit.} Consequently, $|a_i|r_j\to0$ whenever $r_j/L_{\rm AdS}\to0$. The condition \eqref{eq:small-flux}, which supports interpreting the consistent truncation as an effective theory of light modes, is therefore parametrically satisfied in our scale-separated regime.

We now distinguish solutions at weak and strong string coupling.

\subsubsection{Weak string coupling}

The string coupling tends to zero for $\lambda \to \infty$ when
$\gamma<3(\alpha+\beta)$. Hence a large-radius, weakly coupled
regime with scale separation is obtained for
\begin{equation}
    \alpha,\beta>0,
    \qquad
    \alpha+\beta<\gamma<3(\alpha+\beta),
    \qquad
    \frac{\gamma+\alpha+\beta}{4}
    <\delta<
    \frac{\gamma+\alpha+\beta}{2}.
    \label{eq:weak-coupling-window}
\end{equation}
An explicit example is obtained for $\alpha=\beta=1$, $\gamma=4$ and $\delta=2$. The internal radii, string coupling and $L_{\rm AdS}$ then scale as
\begin{equation}
\begin{aligned}
    r_1,r_5,r_4&\sim\lambda^{1/2},
    &r_2,r_3,r_6,r_7&\sim\lambda,\\
    g_s&\sim\lambda^{-1/2},
    &L_{\rm AdS}&\sim\lambda^{3/2}.
\end{aligned}
\end{equation}
The internal volume grows as ${\rm vol}_7\sim\lambda^{11/2}$, while the largest internal radius satisfies $r_{\max}/L_{\rm AdS}\sim\lambda^{-1/2}$. Thus the smeared background admits simultaneous large internal radii, weak string coupling and parametric scale separation.

\subsubsection{Strong string coupling}

The same solution also admits a large-radius, scale-separated
regime with $g_s\to\infty$ as $\lambda \to \infty$, obtained by choosing
\begin{equation}
    \alpha,\beta>0,
    \qquad
    \gamma>3(\alpha+\beta),
    \qquad
    \frac{\gamma+\alpha+\beta}{4}
    <\delta<
    \frac{\gamma+\alpha+\beta}{2}.
    \label{eq:strong-coupling-window}
\end{equation}
An explicit example is obtained for $\alpha=\beta=1$, $\gamma=8$ and $\delta=4$. The internal radii, string coupling and $L_{\rm AdS}$ then scale as
\begin{equation}
\begin{aligned}
    r_1,r_5,r_4&\sim\lambda^{3/2},
    &r_2,r_3,r_6,r_7&\sim\lambda^2,\\
    g_s&\sim\lambda^{1/2},
    &L_{\rm AdS}&\sim\lambda^{5/2}.
\end{aligned}
\end{equation}
The internal volume grows as ${\rm vol}_7\sim\lambda^{25/2}$,
while the largest internal radius satisfies $r_{\max}/L_{\rm AdS}\sim\lambda^{-1/2}$. Thus the smeared background also admits simultaneous large internal radii, strong string coupling and parametric scale separation.

This regime is not perturbatively controlled in type IIA string theory. Since the Romans mass vanishes, it is instead natural to investigate an eleven-dimensional description. A necessary bulk check is that the internal lengths remain large in eleven-dimensional Planck units and that the additional Kaluza-Klein tower associated with the M-theory circle remains above the scale set by the anti-de Sitter radius.
Up to constant numerical factors, the IIA/M-theory dictionary ($\ell_p \simeq g_s^{1/3} \sqrt{\alpha'}$) gives
\begin{equation}
    \frac{\tilde R_i}{\ell_p}\sim\frac{r_i}{g_s^{1/3}},
    \qquad
    \frac{\tilde L_{\rm AdS}}{\ell_p}
        \sim\frac{L_{\rm AdS}}{g_s^{1/3}},
    \qquad
    \frac{R_{11}}{\ell_p}\sim g_s^{2/3},
\end{equation}
where $\tilde R_i$, $i=1,2,\dots,7$, and $\tilde L_{\rm AdS}$ denote the bulk (dimensionful) length scales inherited from the seven-dimensional internal metric, and $R_{11}$ is the M-theory circle radius. In terms of the eleven-dimensional Planck length, $\ell_p$,  we thus have the scalings
\begin{equation}
\begin{aligned}
    \frac{\tilde R_{1}}{\ell_p}\sim     \frac{\tilde R_{5}}{\ell_p}
        &\sim\lambda^{\gamma/6},\\
    \frac{\tilde R_{2}}{\ell_p}\sim     \frac{\tilde R_{6}}{\ell_p}
        &\sim\lambda^{\frac16(3\alpha+\gamma)},\\
    \frac{\tilde R_{3}}{\ell_p}\sim     \frac{\tilde R_{7}}{\ell_p}
        &\sim\lambda^{\frac16(3\beta+\gamma)},\\
    \frac{\tilde R_4}{\ell_p}
        &\sim\lambda^{\delta-\gamma/3},\\
    \frac{R_{11}}{\ell_p}
        &\sim\lambda^{\frac16(\gamma-3\alpha-3\beta)},\\
    \frac{\tilde L_{\rm AdS}}{\ell_p}
        &\sim\lambda^{\frac16(3\alpha+3\beta+\gamma)}.
\end{aligned}
\end{equation}
In the strong-coupling regime, all these internal lengths grow provided we strengthen the last condition on $\delta$ in \eqref{eq:strong-coupling-window} to
\begin{equation}
    \frac{\gamma}{3}<\delta<
    \frac{\gamma+\alpha+\beta}{2}.
    \label{eq:eleven-dimensional-window}
\end{equation}
The ratios of the internal lengths to the anti-de Sitter radius are unchanged, while the M-theory
circle satisfies
\begin{equation}
    \frac{R_{11}}{\tilde L_{\rm AdS}}
    \sim\frac{g_s}{L_{\rm AdS}}
    \sim\lambda^{-\alpha-\beta}
    \to 0,
\end{equation}
thus its tower of Kaluza-Klein excitations also remains parametrically separated from the anti-de Sitter scale.
For the explicit strong-coupling example above, these expressions
reduce to
\begin{equation}
\begin{aligned}
    \tilde R_{1,5,4}/\ell_p&\sim\lambda^{4/3},\qquad\qquad
    &\tilde R_{2,3,6,7}/\ell_p&\sim\lambda^{11/6},\\
    R_{11}/\ell_p&\sim\lambda^{1/3},
    &\tilde L_{\rm AdS}/\ell_p&\sim\lambda^{7/3}.
\end{aligned}
\end{equation}
This explicit family of solutions therefore has the bulk length hierarchies required for a large-radius, scale-separated eleven-dimensional description. Establishing such a description, however, requires further analysis and is left for future work.

\section{Obstruction to $\mathcal{N}=(1,1)$ supersymmetry}
\label{sec:opposite-supersymmetry}

The bispinor equations establish that the solution constructed above preserves at least one three-dimensional supersymmetry. We now ask whether it admits an additional supersymmetry with the opposite sign of the three-dimensional anti-de Sitter Killing spinor equation, leading to an $\mathcal N=(1,1)$ extension.
We first exclude this possibility in the left-invariant sector.
We then extend the argument to coordinate-dependent Killing spinors that are smooth on the covering space. For the unshifted quotient considered here, the orbifold and orientifold involutions have a common fixed point at the origin, where their projections impose the same algebraic restrictions as in the left-invariant sector, leading to the same contradiction. This regularity requirement is appropriate to the smeared background under consideration; we do not extend the conclusion to a localized or resolved background.

\subsection{Spinor conventions and algebraic conditions}

A candidate $\mathcal N=(1,1)$ solution has ten-dimensional supersymmetry parameters of the form
\begin{equation}
\begin{aligned}
    \epsilon_1
    &=\zeta_+\otimes\theta_+\otimes\chi_{1}^+
      +\zeta_-\otimes\theta_+\otimes\chi_{1}^-,\\
    \epsilon_2
    &=\zeta_+\otimes\theta_-\otimes\chi_{2}^+
      +\zeta_-\otimes\theta_-\otimes\chi_{2}^-.
\end{aligned}
\label{eq:epsi12decomp}
\end{equation}
Here, the auxiliary spinors $\theta_\pm$ and seven-dimensional gamma matrices are those introduced above; in particular, the $\gamma_m$ are imaginary and antisymmetric.
The label $\pm$ on $\zeta_\pm$ and $\chi_{1,2}^\pm$ indicate the two three-dimensional supersymmetries. With the dictionary $m=-\mu$
relative to \cite{Macpherson:2021lbr}, the Killing spinor equation in our conventions reads
\begin{equation}
    \nabla_\nu\zeta_\pm
    =\mp\frac{\mu}{2}\gamma^{(3)}_\nu\zeta_\pm,
    \qquad L_{\rm AdS}=|\mu|^{-1}.
\end{equation}
We label the known supersymmetry by $+$ and set $\chi_{1,2}^+\equiv \chi_{1,2}$. On the background under consideration, $A=0$ and $H_{10}=0$.
Due to \eqref{eq:electric-norm}, the vanishing external NS flux implies that the two internal spinors in each AdS$_3$ sector have equal norms, while the constant warp factor makes these norms constant. Rescaling each pair together with its corresponding external spinor $\zeta_\pm$, we may therefore choose $\|\chi_{1}^\pm\|=\|\chi_{2}^\pm\|=1$. Recall that the already established supersymmetry obeys $\chi_1=-i\slashed v\chi_2$. In order to simplify the notation in what follows, let us set $\eta=\chi_{1+}$ and $w=\chi_{2+}$, so that we have (recall that $v$ has unit norm)
\begin{equation}
    w=i\slashed v\eta,
    \qquad \slashed v=v^m\gamma_m,
    \qquad (i\slashed v)^2=-1,
    \qquad \eta^T\eta=w^Tw=1,
    \qquad \eta^Tw=0.
    \label{eq:known-singlet-plane}
\end{equation}
The real spinors $\eta,w$ span the two singlets of the known $\mathrm{SU}(3)$ structure.

Since the warp factor is constant, our metric is a direct product of three-dimensional anti-de Sitter space and $\mathcal M_7$ and cannot locally enhance to a four-dimensional anti-de Sitter factor extending the external directions.\footnote{A four-dimensional anti-de Sitter factor written in terms of three-dimensional anti-de Sitter slices requires a warp factor that varies along the additional direction, whereas the warp factor of our solution is constant.} A necessary condition for the mixed bilinears of a candidate $\mathcal N=(1,1)$ solution is therefore \cite[eq.~(2.9)]{Macpherson:2021lbr}
\begin{equation}
\begin{aligned}
    \eta^T\chi_1^-+w^T\chi_2^-&=0,\\
    \eta^T\gamma_m\chi_1^-
       -w^T\gamma_m\chi_2^-&=0,
\end{aligned}
    \label{eq:mixed-AdS3-conditions}
\end{equation}
with signs chosen for type IIA.

\subsection{The left-invariant sector}
The orbifold generators \eqref{eq:OrbifoldGen} act on the orthonormal coframe as
\begin{equation}
    \Theta_a^*e^i=s_i^{(a)}e^i,
    \qquad s_i^{(a)}=\pm1,
\end{equation}
with $a=1,2$, $i=1,\dots,7$ and no summation over $i$.
Their action on spinors is specified by spin lifts $\widehat\Theta_a$, which implement the same transformation on the gamma matrices:
\begin{equation}
    \widehat\Theta_a\gamma_i\widehat\Theta_a^{-1}
    =s_i^{(a)}\gamma_i.
\end{equation}
These lifts are part of the definition of the orbifold and must obey its group relations. We use those preserving the already established supersymmetry, so that $\widehat\Theta_a\eta=\eta$. Consequently,
\begin{equation}
    \widehat\Theta_a(\gamma_i\eta)
    =\bigl(\widehat\Theta_a\gamma_i\widehat\Theta_a^{-1}\bigr)
      \widehat\Theta_a\eta
    =s_i^{(a)}\gamma_i\eta,
    \label{eq:Thetaetai}
\end{equation}
implying that the spinors $\gamma_i\eta$ have the same orbifold parities as the corresponding one-forms $e^i$. In particular, since $v=e^4$ is invariant under both orbifold generators, $w=i\slashed v\eta$ is also invariant. Any putative additional supersymmetry must be invariant under the same spin lifts.

Since $\eta$ is a unit spinor, the eight real spinors $\eta,i\gamma_i\eta$ form an orthonormal basis. We first expand $\chi_1^-$ in this basis; the same argument applies to $\chi_2^-$, which will subsequently be determined by the orientifold projection.
A general left-invariant candidate therefore takes the form
\begin{equation}
    \chi_1^-=c_0\eta+i\sum_{i=1}^7c_i\,\gamma_i\eta\,,
    \qquad c_0^2+\sum_{i=1}^7c_i^2=1 ,
    \label{eq:chi1m1}
\end{equation}
where $c_0$, $c_i$ are constants. Due to \eqref{eq:Thetaetai}, invariance of $\chi_1^-$ under the orbifold generators requires $c_i=0$ whenever
$s_i^{(a)}=-1$ for any $a=1,2$.
The parities given in \eqref{eq:OrbifoldGen} imply that
$\Theta_1$ sets $c_1,c_2,c_5,c_6$ to zero, while $\Theta_2$
sets $c_1,c_3,c_5,c_7$ to zero. Only $c_0$ and $c_4$
can therefore survive, giving
\begin{equation}
    \chi_1^-=c_0\eta+c_4w,
    \qquad c_0^2+c_4^2=1.
    \label{eq:opposite-singlet-spinor}
\end{equation}
The same restriction applies to $\chi_2^-$. Thus all four internal spinors lie in the singlet space of the same $\mathrm{SU}(3)$ structure as the already established solution.\footnote{If $c_4\neq0$, $\eta$ and $\chi_1^-$ are independent and define the same $\mathrm{SU}(3)$ structure as $\eta,w$. If $c_4=0$, they are parallel, but the known pair $\eta,w$ still fixes the common structure to $\mathrm{SU}(3)$.} An additional left-invariant supersymmetry therefore cannot reduce their common stabilizer further, for example to $\mathrm{SU}(2)$, since the orbifold projections remove the spinor components required for such a reduction. It remains to test the surviving $\mathrm{SU}(3)$ candidates against the orientifold projection and the necessary conditions \eqref{eq:mixed-AdS3-conditions}.

We now impose the orientifold projection, following the discussion and notation around \eqref{eq:e1Ge2}. It is sufficient to consider $O6_\alpha$, which wraps the internal directions $2345$, since the remaining O6 involutions differ from it by orbifold generators.
Its projection relates the ten-dimensional supersymmetry
parameters as
\begin{equation}
    \epsilon_1
    =\sigma_{2345}\Gamma_{\mathrm{AdS}_3}
      \Gamma_{2345}\epsilon_2.
\end{equation}
Upon inserting the spinor decomposition \eqref{eq:epsi12decomp}, this gives
\begin{equation}
    \chi_1^\pm=\sigma\,\gamma_{2345}\chi_2^\pm,
    \qquad \sigma=\pm1,
    \label{eq:internal-O6-projection}
\end{equation}
where $\sigma$ includes the source orientation and the sign coming from the external gamma-matrix product. Since $(\gamma_{2345})^2=1$, the already established pair $\chi_1^+=\eta$, $\chi_2^+=w$ satisfies $\sigma\gamma_{2345}w=\eta$ and $ \sigma\gamma_{2345}\eta=w$.
Applying the same projection to the candidate additional supersymmetry therefore determines $\chi_2^-$ as
\begin{equation}
    \chi_2^-=\sigma\gamma_{2345}\chi_1^-
    =\sigma\gamma_{2345}(c_0\eta+c_4w)
    =c_4\eta+c_0w.
\end{equation}
Substitution into the first of necessary conditions in \eqref{eq:mixed-AdS3-conditions} and using \eqref{eq:known-singlet-plane} leads to $2c_0=0$. Contracting with $v^m$ the second of the necessary conditions in \eqref{eq:mixed-AdS3-conditions},  and using $\slashed v\eta=-iw$, $\slashed v w=i\eta$ and $\eta^T \gamma_m \eta=0$, gives
\begin{equation}
    \eta^T\slashed v(c_0\eta+c_4w)
    -w^T\slashed v(c_4\eta+c_0w)
    =2ic_4=0.
\end{equation}
Hence we arrived at $c_0=c_4=0$, contradicting $c_0^2+c_4^2=1$. There is therefore no left-invariant $\mathcal N=(1,1)$ extension compatible with the chosen projections.

\subsection{Extension to coordinate-dependent spinors}

The previous conclusion relies on the coefficients being constant. For coordinate-dependent internal spinors, the expansion becomes
\begin{equation}
    \chi_1^-(y)
    =c_0(y)\eta(y)
      +i\sum_{i=1}^7c_i(y)\,\gamma_i\eta(y).
\end{equation}
Orbifold invariance now imposes parity conditions on the coefficient functions:
\begin{equation}
    c_0(\Theta_a y)=c_0(y),
    \qquad
    c_i(\Theta_a y)=s_i^{(a)}c_i(y),
    \qquad a=1,2,
\end{equation}
with no summation over $i$.
Unlike the constant coefficients considered after \eqref{eq:chi1m1}, odd functions need not vanish identically. The orbifold projections therefore do not restrict the spinors to the singlet space of the known $\mathrm{SU}(3)$ structure at a generic point.

The preceding no-go nevertheless extends to spinors that are smooth on the covering space at a common fixed point $y_*$ of the orbifold and orientifold involutions. For the quotient without shifts considered here, the origin is such a point. Since $\Theta_a y_*=y_*$, the parity conditions become
\begin{equation}
    c_i(y_*)=s_i^{(a)}c_i(y_*).
\end{equation}
Every coefficient odd under either generator must therefore
vanish there, leaving only
\begin{equation}
    \chi_1^-(y_*)
    =c_0(y_*)\eta(y_*)+c_4(y_*)w(y_*).
\end{equation}
The same restriction applies to $\chi_2^-(y_*)$, allowing
us to repeat the algebraic argument of the left-invariant
sector at this point.

Away from the fixed locus, the orientifold projection relates
spinors at points exchanged by the involution. At $y_*$,
these points coincide, and the projection takes the pointwise
form \eqref{eq:internal-O6-projection}. Consequently,
\begin{equation}
    \chi_2^-(y_*)
    =\sigma\gamma_{2345}\chi_1^-(y_*)
    =c_4(y_*)\eta(y_*)+c_0(y_*)w(y_*).
\end{equation}
The same substitution into \eqref{eq:mixed-AdS3-conditions} as in the left-invariant sector then gives $c_0(y_*)=0=c_4(y_*)$. Hence $\chi_1^-(y_*)=0$, contradicting its constant unit norm. 
We therefore also exclude $\mathcal N=(1,1)$ enhancements with coordinate-dependent internal Killing spinors that are smooth on the covering space at the common fixed point.

This conclusion relies on the existence of a common fixed point and the regularity of the spinors there. If shifts in the quotient remove the common fixed point, or if the spinors do not admit a smooth covering-space description at that locus, this argument does not apply. In particular, extending the conclusion to a resolved background or one with fully localized sources requires a separate analysis.
Finally, excluding $\mathcal N=(1,1)$ does not forbid an $\mathcal N=(2,0)$ extension.

\section{Gauged supergravity and partial supersymmetry breaking}
\label{sec:3d-effective-theory}

We now turn to the three-dimensional effective description of the solutions constructed above. We first consider a holomorphic superpotential that reproduces the scalar potential at zero axions and its supersymmetric anti-de Sitter vacuum. However, interpreting this superpotential within an ungauged supergravity theory leads to a puzzle: the vacuum would be $\mathcal{N}=(1,1)$, in contradiction with the obstructions of section~\ref{sec:opposite-supersymmetry}. Motivated by this, we then include the axionic gauging required by the reduction and show that the full bosonic action is reproduced precisely by a gauged $\mathcal N=2$ supergravity. Crucially, the gauging leads to a spontaneous breaking of supersymmetry such that only $\mathcal{N}=1$ is preserved on the vacuum.\footnote{The presence of at least $\mathcal{N}=1$ preserved supersymmetry on the vacuum is indeed verified in section \ref{sec:SU3solution} by solving the bispinor equations.} The dimensional reduction is performed in detail in appendix~\ref{app:dimensional-reduction}; here we concentrate on its supergravity interpretation.

\subsection{Fields and conventions}

We describe the scalar sector in terms of eight
three-dimensional $\mathcal N=2$ chiral multiplets, whose
complex scalars are
\begin{equation}
 X^A=x_A+i\xi_A,\qquad
 T_a=t_a+i b_a,\qquad
 Y=y+i\upsilon,
\end{equation}
with $A=0,1,2,3$ and $a=1,2,3$. We denote $X^A=(S,U_1,U_2,U_3)$, whose real parts are related to the internal radii and dilaton by \eqref{eq:dr-scalar-dictionary}, while the axions are defined by \eqref{eq:dr-axion-conventions}.
The description of $Y$ uses the first-order formulation of appendix~\ref{app:dr-CS}, in which the scalar $\upsilon$ is introduced together with the auxiliary vector $\mathcal B$, without adding physical degrees of freedom.

Denoting these complex scalars collectively by $Z^I=(X^A,T_a,Y)$, their kinetic terms determine the K\"ahler metric $g_{I\bar J}=\partial_I\partial_{\bar J}K$,
with
\begin{equation}
 K=-\log(Y+\bar Y)
   -\sum_{A=0}^3\log(X^A+\bar X^A)
   -\sum_{a=1}^3\log(T_a+\bar T_a).
 \label{eq:eft-K}
\end{equation}
This reproduces the scalar kinetic terms in \eqref{eq:dr-Maxwell-description}, supplemented by the $\upsilon$ term in \eqref{eq:dr-first-order}. The dictionary \eqref{eq:dr-scalar-dictionary} also gives the useful relations
\begin{equation}
 \prod_Ax_A=\frac{\mathcal V_{6,S}^2}{g_s^4},
 \qquad
 e^K=\frac{g_s^6}{256r_4^2\mathcal V_{6,S}^4}
     =\frac{e^{4D}g_s^2}{256\mathcal V_{6,S}^2}.
 \label{eq:eft-identities}
\end{equation}
Throughout this section, we use the three-dimensional Einstein-frame normalization of \eqref{eq:dr-3d-action}, with gravitational term $R_{3,E}/2$. The potential at an anti-de Sitter vacuum therefore satisfies $V_0=-L_{3,E}^{-2}$, where $L_{3,E}$ is related to the string frame radius $L_{\rm AdS}$ by \eqref{eq:dr-AdS-radius}.

\subsection{An apparent superpotential description}

We first ask whether the solution can be described by an $\mathcal N=2$ theory whose scalar potential is determined entirely by the K\"ahler potential and a holomorphic superpotential. In the absence of a gauging contribution, the full scalar potential takes the form
\begin{equation}
 V_F[W]=e^K\left(
 g^{I\bar J}D_IW D_{\bar J}\bar W-4|W|^2
 \right),
 \qquad
 D_IW=\partial_IW+K_IW.
 \label{eq:eft-VFdefinition}
\end{equation}
We will first construct a candidate holomorphic superpotential that reproduces the dimensionally reduced potential \eqref{eq:dr-compact-potential} at zero axions and the ten-dimensional solution \eqref{eq:flux-radius-relations}.
We will then show that this description predicts an amount of preserved supersymmetry on the vacuum incompatible with the analysis of section~\ref{sec:opposite-supersymmetry}.

Using the flux coefficients introduced in \eqref{RR2_expansion}--\eqref{RR6_expansion}, we begin with the superpotential for all scalars but $Y$, which reads \cite{Camara:2005dc}
\begin{equation}
 W_0=4\left[
 hT_1\sum_AX^A+M+NT_1T_3+R_FT_1T_2+m_0T_2T_3
 \right].
 \label{eq:eft-W0}
\end{equation}
This expression includes the metric flux and the two-form and six-form RR fluxes. To incorporate the background four-form contribution, as well as the dependence on $Y$ (parametrizing the circle) we denote
\begin{equation}
 (Q_0,Q_1,Q_2,Q_3)=(f_3,f_1,f_2,f_4),
 \qquad \text{such that} \qquad  \sum_AQ_A=0,
 \label{eq:eft-charges}
\end{equation}
where the last equality is the Bianchi identity
\eqref{eq:dF4tad}. We then consider the holomorphic
superpotential
\begin{equation}
 W=W_0+H,
 \label{eq:eft-auxW}
\end{equation}
with
\begin{equation}
 H=-4\left(\sum_A\frac{Q_A}{X^A}\right)
 \sqrt{\frac{X^0X^1X^2X^3T_1T_2T_3}{Y}}. 
\end{equation}

At zero axions, substituting \eqref{eq:eft-auxW} into
\eqref{eq:eft-VFdefinition} gives
\begin{equation}
 V_F[W]\big|_{\text{axions}=0}
 =V_{\mathrm{no\,axions}}
 +\frac{e^{5\phi}ht_1}{2\mathcal V_{7,S}^3}
   (f_1+f_2+f_3+f_4),
\end{equation}
where $V_{\mathrm{no\,axions}}$ is the potential obtained by direct dimensional reduction in \eqref{eq:dr-compact-potential} and $\mathcal{V}_{7,S}$ is given in \eqref{eq:V6sV7s}. The last term vanishes upon imposing the Bianchi identity \eqref{eq:dF4tad}, equivalently $\sum_AQ_A=0$, and the result can be written as
\begin{equation}
 V_F[W]\big|_{\text{axions}=0}
 =V_F[W_0]\big|_{\text{axions}=0}
 +\frac1{4y^2}\sum_A\frac{Q_A^2}{x_A^2}
 =V_{\mathrm{no\,axions}}.
 \label{eq:eft-auxmatch}
\end{equation}
Thus $H$ supplies the background four-form energy \eqref{eq:dr-F4E}, while $W_0$ reproduces the remaining contributions, with the matching holding for arbitrary real saxions.
Moreover, evaluating on the solution \eqref{eq:flux-radius-relations} gives
\begin{equation}
 D_IW\big|_*=0,\qquad
 \text{and}\qquad
 V_F[W]\big|_*
 =-4e^K W \bar W\big|_*=-\frac{e^{4D_0}\Delta^2}{36}
 =-\frac1{L_{3,E}^2}.
 \label{eq:eft-aux-vacuum}
\end{equation}
The ten-dimensional solution appears thus as a supersymmetric anti-de Sitter vacuum of this description, with the correct cosmological constant.

This apparent agreement nevertheless leads to an incorrect supersymmetry counting. In an ungauged $\mathcal N=2$ theory, an anti-de Sitter vacuum satisfying the F-term equations preserves both supersymmetries. Indeed, $D_IW=0$ make all matter-fermion shifts vanish, while the gravitino
variations take the form\footnote{The external gamma matrices used in this section are related to those entering the ten-dimensional spinor decomposition by $\gamma_{\nu,E}=-e^{-2D}\gamma^{(3)}_\nu$ on the vacuum.
Consequently, the ten-dimensional equation $\nabla_\nu\zeta_+=-\frac{\mu}{2}\gamma^{(3)}_\nu\zeta_+$ becomes $ \nabla_\nu\zeta_+  =\frac{e^{2D}\mu}{2}\gamma_{\nu,E}\zeta_+  =-\mathcal P\big|_*\,\gamma_{\nu,E}\zeta_+$, where $\mu=-\Delta/6$ and $\mathcal P|_*=e^{2D}\Delta/12$. This agrees with the preserved $r=2$ gravitino equation in section \ref{sec:3dvacuum}.}
\begin{equation}
 \delta\psi_\mu^r
 =\nabla_\mu\epsilon^r+(A_1)^{rs}\gamma_\mu\epsilon^s,
 \qquad r,s=1,2,
\end{equation}
where $\gamma_\mu$ refers to the three-dimensional
Einstein-frame metric. With no gauging contribution, the shift matrix $A_1$ is given by \cite{deWit:2003ja}
\begin{equation}
 A_1=e^{K/2}
 \begin{pmatrix}
 -\mathrm{Re}\,W
 &\mathrm{Im}\,W\\
 \mathrm{Im}\,W
 &\mathrm{Re}\,W
 \end{pmatrix}.
\end{equation}
On the vacuum, $W$ is real and
\eqref{eq:eft-aux-vacuum} gives
\begin{equation}
 A_1\big|_*
 =\frac{e^{2D}\Delta}{12}
 \begin{pmatrix}
 -1&0\\
 0&1
 \end{pmatrix},
\end{equation}
with eigenvalues $\pm(2L_{3,E})^{-1}$. Denoting the corresponding supersymmetry parameters by $\epsilon_\pm$, the vanishing gravitino variations give 
\begin{equation}
 \nabla_\mu\epsilon_+
 =-\frac{1}{2L_{3,E}}\gamma_\mu\epsilon_+,
 \qquad
 \nabla_\mu\epsilon_-
 =+\frac{1}{2L_{3,E}}\gamma_\mu\epsilon_-,
\end{equation}
implying that both three-dimensional Killing-spinor sectors are preserved, so that this description predicts an $\mathcal N=(1,1)$ vacuum. This contradicts the obstruction established in section~\ref{sec:opposite-supersymmetry}.

The identity \eqref{eq:eft-auxmatch} involving the scalar potential is therefore insufficient to identify the effective theory: it tests only the scalar potential at zero axions and does not account for the axionic gauge couplings derived in the appendix. We now include these couplings and explain both how the gauged description resolves the discrepancy and why the $W$ introduced above reproduces the real-moduli potential despite failing to describe the full theory.

Note that we do not consider altering the $W_0$ part of the superpotential or reinterpreting it as a gauging because it is in some sense directly inherited in this form from the known four-dimensional constructions \cite{Camara:2005dc}, therefore our main concern is to properly describe the effect of reducing further the four-dimensional theory on the circle. Of course to capture properly the reduction we had to work it out in the full ten-dimensional theory as we have been doing here.

\subsection{Matching with gauged supergravity}

We now include the gauge couplings obtained by dimensional reduction. As shown in \eqref{eq:dr-axion-gauging}, the Kaluza-Klein vector $\mathcal A$ associated to the circle gauges the RR axion shifts through $D\xi_A=d\xi_A-Q_A\mathcal A$. In the equivalent first-order formulation of appendix~\ref{app:dr-CS}, the auxiliary vector $\mathcal B$ also gauges the shift of $\upsilon$. In terms of the complex scalar coordinates, these couplings take the form
\begin{equation}
 DX^A=dX^A-iQ_A\mathcal A,\qquad
 DY=dY-i\mathcal B,\qquad
 DT_a=dT_a,
 \label{eq:eft-covariant}
\end{equation}
together with the Chern--Simons term
\begin{equation}
 S_{\mathrm{CS}}
 =\frac1{2\kappa_3^2}\int\mathcal B\wedge d\mathcal A.
 \label{eq:eft-CS}
\end{equation}

To express these couplings in the language of gauged
supergravity, we introduce the Killing vectors generating
the axionic shifts,
\begin{equation}
 k_Y=i(\partial_Y-\partial_{\bar Y}),\qquad
 k_A=i(\partial_{X^A}-\partial_{\bar X^A}),\qquad
 k_Q=\sum_AQ_Ak_A.
\end{equation}
The covariant derivatives \eqref{eq:eft-covariant} show that $\mathcal A$ gauges $k_Q$, while $\mathcal B$ gauges $k_Y$. These Killing vectors generate independent axionic translations and therefore commute. Since the K\"ahler potential depends only on the real scalar components, it is invariant under these shifts.
Each Killing vector is associated with a real function $\mu_M$, called its moment map. Denoting the K\"ahler form of the scalar manifold by $\mathcal K=i g_{I\bar J}dZ^I\wedge d\bar Z^{\bar J}$, we define these functions through $\iota_{k_M}\mathcal K=d\mu_M$; for example, \eqref{eq:eft-K} gives $ \iota_{k_Y}\mathcal K  =-\frac{dY+d\bar Y}{(Y+\bar Y)^2} =d\left(\frac1{Y+\bar Y}\right)$.
The moment maps associated to our gaugings are therefore
\begin{equation}
 \mu_Y=\frac1{Y+\bar Y},\qquad
 \mu_A=\frac1{X^A+\bar X^A},
 \label{eq:eft-moment-maps}
\end{equation}
where we have already set to zero the possible additive constants, usually called Fayet--Iliopoulos parameters.

The embedding tensor $\Theta_{MN}$ specifies which combinations of the isometries on the scalar-manifold are coupled to the vectors.
Following the construction of \cite[eqs.~(3.3), (3.10)]{deWit:2003ja}, we  introduce it by setting
\begin{equation}
 DZ^I\equiv dZ^I-\Theta_{MN}\mathscr A^M k_N^I,
 \qquad
 S_{\mathrm{CS}}
 \equiv \frac1{4\kappa_3^2}\int
 \Theta_{MN}\mathscr A^M\wedge d\mathscr A^N,
 \label{eq:eft-CS-convention}
\end{equation}
where $M,N$ label the generators $(k_Y,k_A)$ and $k_M^I=k_M(Z^I)$ are their holomorphic components. Thus, each vector $\mathscr A^M$ couples to the combination $\Theta_{MN}k_N$, and the same symmetric tensor determines the Chern--Simons couplings.
The covariant derivatives \eqref{eq:eft-covariant}
and the Chern--Simons coupling \eqref{eq:eft-CS},
obtained by dimensional reduction in
appendix~\ref{app:dimensional-reduction}, are reproduced by
\begin{equation}
 \Theta_{YA}=\Theta_{AY}=Q_A,\qquad
 \Theta_{YY}=\Theta_{AB}=0,\qquad
 \mathscr A^Y=\mathcal A,\qquad
 \sum_AQ_A\mathscr A^A=\mathcal B.
 \label{eq:eft-embedding}
\end{equation}
This embedding tensor is symmetric and satisfies the quadratic constraint expressing its gauge invariance \cite[eq.~(3.13)]{deWit:2003ja}. Indeed, all of our Killing vectors generate commuting axionic shifts, so the structure constants vanish. The moment maps are also compatible with the abelian gauge algebra: they depend only on $y$ and $x_A$, so they are invariant under both gauged axionic shifts. Thus $k_M(\mu_N)=0$, as required by equivariance.
Notice that, although the notation $\mathscr A^M$ introduces five vector labels, only $\mathcal A$ and $\mathcal B=\sum_AQ_A\mathscr A^A$ enter the action. For charges $Q_A$ not all zero, the embedding tensor has rank two, consistently with the fact that $\mathcal A$ gauges $k_Q=\sum_AQ_Ak_A$, while $\mathcal B$ gauges $k_Y$. The remaining three combinations of $\mathscr A^A$ lie in the kernel of $\Theta_{MN}$ and do not play any active role. Thus, this formalism describes precisely the two commuting gauged shifts already identified in the reduced action, without introducing additional physical vectors.

The gauging contribution to the scalar potential is governed by a real function quadratic in the moment maps. In \cite{deWit:2003ja}, this function is called $T$-tensor, denoted by $T$, and is obtained by contracting the moment maps with the embedding tensor. To avoid confusion with the chiral fields $T_a$, we introduce the notation
\begin{equation}
 \mathcal P=\frac12\Theta_{MN}\mu_M\mu_N
 =\frac1{Y+\bar Y}\sum_A\frac{Q_A}{X^A+\bar X^A},
 \label{eq:eft-P}
\end{equation}
where the second equality follows from \eqref{eq:eft-moment-maps} and \eqref{eq:eft-embedding}.\footnote{Our definition $\iota_{k_M}\mathcal K=d\mu_M$ gives $\partial_I\mu_M=-i g_{I\bar J}k_M^{\bar J}$, whereas \cite[eq.~(2.30)]{deWit:2003ja} uses $\partial_I\mathcal V^{M\,12}=\frac{i}{2}g_{I\bar J}k_M^{\bar J}$. Thus, we identify $\mu_M=-2\mathcal V^{M\,12}$ for the same Killing vectors and vanishing additive constants. We also absorb their gauge coupling into the embedding tensor, $\Theta_{MN}^{\mathrm{here}} =g\Theta_{MN}^{\mathrm{there}}$. Combining the definitions $T^{12,12}=\Theta_{MN}^{\mathrm{there}} \mathcal V^{M\,12}\mathcal V^{N\,12}$ and $T=2T^{12,12}$ in equations~(3.14) and (4.4) of \cite{deWit:2003ja}, we obtain $\mathcal P=gT=2\Theta_{MN}^{\mathrm{here}}(-\mu_M/2)(-\mu_N/2) =\frac12\Theta_{MN}^{\mathrm{here}}\mu_M\mu_N$.}

We retain $W_0$ in \eqref{eq:eft-W0} as our holomorphic superpotential and check its compatibility with the gauging. Since the K\"ahler potential is invariant and we have introduced no additional constant R-symmetry contributions, compatibility of the superpotential with the gauging means that $W_0$ must be invariant under both gauged generators. We can check that this is the case, indeed
\begin{equation}
 k_YW_0=0,\qquad
 k_QW_0
 =i\sum_AQ_A\partial_{X^A}W_0
 =4ihT_1\sum_AQ_A=0.
\end{equation}
The first equality follows because $W_0$ is independent of $Y$, while the second uses the four-form Bianchi identity \eqref{eq:dF4tad}, rewritten as in \eqref{eq:eft-charges}. Thus, the complete potential of the gauged $\mathcal N=2$ supergravity reads \cite{deWit:2003ja}
\begin{equation}
 V=V_F[W_0] +4\left( g^{I\bar J}\partial_I\mathcal P \partial_{\bar J}\mathcal P-\mathcal P^2 \right).
 \label{eq:eft-V}
\end{equation}

We now show how this supergravity potential reproduces exactly the one obtained in appendix \ref{app:dimensional-reduction} by direct dimensional reduction.  Since $\mathcal P$ depends only on the real parts of the fields, we have
\begin{equation}
 \mathcal P=\frac1{4y}\sum_A\frac{Q_A}{x_A},\qquad
 \partial_{X^A}\mathcal P=-\frac{Q_A}{8yx_A^2},
 \qquad
 \partial_Y\mathcal P=-\frac{\mathcal P}{2y},
 \qquad
 \partial_{T_a}\mathcal P=0.
\end{equation}
One can then check that
\begin{equation}
 4\left( g^{I\bar J}\partial_I\mathcal P   \partial_{\bar J}\mathcal P-\mathcal P^2\right) =\frac1{4y^2}\sum_A\frac{Q_A^2}{x_A^2} =V_{F_4},
 \label{eq:eft-F4energy}
\end{equation}
thus the gauging reproduces exactly the four-form potential \eqref{eq:dr-F4E}.

For the remaining contributions, substituting $K$ and
$W_0$ into \eqref{eq:eft-VFdefinition} gives
\begin{equation}
\begin{aligned}
 V_F[W_0]={}&
 \underbrace{64e^K h^2t_1^2\sum_Ax_A^2}_{V_R}
 +\underbrace{64e^K h^2b_1^2\sum_Ax_A^2}_{V_{H_3}}
 \\
 &+\underbrace{64e^K\left(
 m_0^2t_2^2t_3^2+N^2t_1^2t_3^2
 +R_F^2t_1^2t_2^2\right)}_{V_{F_2}}
 \\
 &+\underbrace{64e^K\left(
 t_1^2\mathcal R_{23}^2+t_2^2\mathcal R_{13}^2
 +t_3^2\mathcal R_{12}^2\right)}_{V_{\mathcal G_4}}
 \\
 &+\underbrace{64e^K(M+\rho)^2}_{V_{\mathrm{el}}}
 -\underbrace{128e^Khm_0\mathcal V_{6,S}
 \sum_Ax_A}_{-V_{\mathrm{O6}}}.
 \label{eq:eft-VFexpanded}
\end{aligned}
\end{equation}
Here, $\mathcal R_{ab}$ and $\rho$ are defined in \eqref{eq:dr-axion-polynomials} and \eqref{eq:dr-electric-polynomial}. The identifications under each term follow from \eqref{eq:dr-scalar-dictionary} and \eqref{eq:eft-identities}, and reproduce the curvature, internal flux, electric and source contributions derived in appendix~\ref{app:dr-potential}.
We therefore obtain
\begin{equation}
V_F[W_0]  +4\left( g^{I\bar J}\partial_I\mathcal P \partial_{\bar J}\mathcal P-\mathcal P^2 \right)= V_{\mathrm{full}},
 \label{eq:eft-full-potential-match}
\end{equation}
where $V_{\mathrm{full}}$ is the potential derived in \eqref{eq:dr-full-axion-potential}, including axions. Together with the K\"ahler metric, covariant derivatives and BF coupling, it establishes the matching of the (bosonic) action.

Finally, we explain why the auxiliary superpotential
$W=W_0+H$ reproduces the scalar potential at zero axions, despite failing to describe the effective theory microscopically. In the gauged theory, the four-form flux contribution is generated by $\mathcal P$, whereas in the auxiliary description it is generated by the holomorphic term $H$.
These functions play different roles, but their contributions to the potential coincide for vanishing axions.
To see this, we expand the F-term potential as
\begin{equation}
\begin{split}
 V_F[W_0+H]
 ={}&V_F[W_0]+V_F[H]+2e^K\mathrm{Re}\left(
 g^{I\bar J}D_IW_0D_{\bar J}\bar H
 -4W_0\bar H\right).
\end{split}
\label{eq:VFspliWH}
\end{equation}
At zero axions, the explicit expressions for $H$ and $\mathcal P$ give
\begin{equation}
 e^{K/2}H\big|_{\text{axions}=0}
 =-\mathcal P\big|_{\text{axions}=0},
 \qquad
 e^{K/2}D_IH\big|_{\text{axions}=0}
 =-2\partial_I\mathcal P\big|_{\text{axions}=0},
 \label{eq:eft-auxidentities}
\end{equation}
as well as 
\begin{equation}
 g^{I\bar J}D_IW_0\,\partial_{\bar J}\mathcal P
 \big|_{\text{axions}=0}
 =
 \left[
 2\mathcal P W_0-\frac{2ht_1}{y}\sum_AQ_A
 \right]_{\text{axions}=0}
 =2\mathcal P W_0\big|_{\text{axions}=0},
 \label{eq:eft-crossidentity}
\end{equation}
where the last equality uses the four-form Bianchi identity. As can be seen from \eqref{eq:eft-auxmatch} and \eqref{eq:eft-F4energy}, the terms quadratic in $H$ reproduce precisely the gauging contribution,
\begin{equation}
 V_F[H]\big|_{\text{axions}=0}
 =4\left(
 g^{I\bar J}\partial_I\mathcal P
 \partial_{\bar J}\mathcal P-\mathcal P^2
 \right)\big|_{\text{axions}=0},
\end{equation}
while the mixed terms, namely the last parenthesis in \eqref{eq:VFspliWH}, vanish due to \eqref{eq:eft-crossidentity}.  Consequently, we established that
\begin{equation}
 V_F[W]\big|_{\text{axions}=0}
 =\left[
 V_F[W_0]
 +4\left(
 g^{I\bar J}\partial_I\mathcal P
 \partial_{\bar J}\mathcal P-\mathcal P^2
 \right)\right]_{\text{axions}=0},
 \label{eq:eft-aux-gauged-identity}
\end{equation}
explaining the agreement in \eqref{eq:eft-auxmatch}.
However, $H$ cannot be included in the superpotential
of the gauged theory because it is not invariant under
the gauged shift of $\upsilon$. Indeed, its dependence
on $Y$ gives
\begin{equation}
 k_YH=i\partial_YH=-\frac{i}{2Y}H,
\end{equation}
which does not vanish identically. Since $k_YW_0=0$, the same shift changes $W=W_0+H$. Consequently, $W$ is incompatible with the gauge couplings obtained by dimensional reduction.

\subsection{The vacuum and partial supersymmetry breaking}
\label{sec:3dvacuum}

We now determine the supersymmetric anti-de Sitter vacua of the gauged theory and show that the ten-dimensional solution \eqref{eq:flux-radius-relations} satisfies their defining conditions. Supersymmetry requires the vanishing of the gravitino and matter-fermion variations for at least one nonzero supersymmetry parameter. At constant scalar fields and vanishing vector field strengths, these conditions take the form \cite{deWit:2003ja}
\begin{equation}
 \nabla_\mu\epsilon^r
 +(A_1)^{rs}\gamma_\mu\epsilon^s=0,
 \qquad
 (A_2)_I^{rs}\epsilon^r=0, \qquad r,s=1,2.
\end{equation}
Here, $A_1$ and $A_2$ denote the gravitino and spin-$1/2$ shifts, respectively.
In our conventions, the gravitino-shift matrix is
\begin{equation}
 A_1=
 \begin{pmatrix}
 -\mathcal P-e^{K/2}\mathrm{Re}\,W_0
 &e^{K/2}\mathrm{Im}\,W_0\\
 e^{K/2}\mathrm{Im}\,W_0
 &-\mathcal P+e^{K/2}\mathrm{Re}\,W_0
 \end{pmatrix},
 \label{eq:eft-A1}
\end{equation}
while the independent holomorphic components of the matter-fermion shifts are\footnote{To fix the relative normalization, we use \cite[eq.~(3.21)]{deWit:2003ja}. For $\mathcal N=2$, its diagonal components give $(A_2)_I^{rr}=\frac12\mathcal D_I(A_1)^{rr}$, without summing over $r$, and with no $T$-tensor contribution because $T_I^{rs}$ is antisymmetric in $r,s$. The derivative includes the $SO(2)$ connection $Q_I^{12}=-iK_I/4$, so that, for example, $\mathcal D_I(A_1)^{11} =\partial_I(A_1)^{11}+2Q_I^{12}(A_1)^{12}$. Using $\partial_I\mathrm{Re}\,W_0 =\frac12\partial_IW_0$, which follows from holomorphicity, we obtain $\mathcal D_I(A_1)^{11} =-\partial_I\mathcal P -\frac12e^{K/2}(\partial_IW_0+K_IW_0)$. The connection term supplies the imaginary part needed to combine $\mathrm{Re}\,W_0$ and $\mathrm{Im}\,W_0$ into $W_0$. The same calculation for $(A_1)^{22}$ gives $-\partial_I\mathcal P+\frac12e^{K/2}D_IW_0$. Multiplying by the prefactor $1/2$ then yields \eqref{eq:eft-A2}. The resulting superpotential coefficient differs by a factor of two from equation~(4.21) as printed in the arXiv version
of that reference.}
\begin{equation}
 (A_2)_I^{11}
 =-\frac14\left(
 e^{K/2}D_IW_0+2\partial_I\mathcal P
 \right),
 \qquad
 (A_2)_I^{22}
 =\frac14\left(
 e^{K/2}D_IW_0-2\partial_I\mathcal P
 \right),
 \label{eq:eft-A2}
\end{equation} 
and the remaining entries satisfy $(A_2)_I^{12}=i(A_2)_I^{11}$ and $(A_2)_I^{21}=-i(A_2)_I^{22}$ \cite{deWit:2003ja}.

The ten-dimensional solution has vanishing axions, so $W_0$ is real and $A_1$ is diagonal, allowing us to test the two supersymmetry sectors separately. Since the solution preserves at least $\mathcal N=1$ supersymmetry, we first identify the preserved supersymmetry in the three-dimensional theory and then show that the other one is broken, thus establishing partial supersymmetry breaking. We label the candidate preserved sector by $r=2$: the vanishing of its spin-$\tfrac12$ variations imposes $(A_2)_I^{22}=0$ for every $I$, while its gravitino variation must reproduce a three-dimensional Killing-spinor equation.
Using \eqref{eq:eft-A1} and \eqref{eq:eft-A2}, these requirements become
\begin{equation}
 e^{K/2}D_IW_0=2\partial_I\mathcal P,
 \qquad
 \left|e^{K/2}W_0-\mathcal P\right|
 =\frac1{2L_{3,E}}.
 \label{eq:eft-susy-conditions}
\end{equation}
The $r=1$ sector would instead require $e^{K/2}D_IW_0=-2\partial_I\mathcal P$ and $|e^{K/2}W_0+\mathcal P|=1/(2L_{3,E})$.

We now verify \eqref{eq:eft-susy-conditions} on the ten-dimensional solution \eqref{eq:flux-radius-relations}, taking $h,m_0>0$ for definiteness.  Using the scalar definitions \eqref{eq:dr-scalar-dictionary}, the charge assignment \eqref{eq:eft-charges}, and the
four-form flux relations in \eqref{eq:flux-radius-relations}, we obtain $\sum_A\frac{Q_A}{x_A}=r_4\frac{\Delta}{3}$. Substituting into \eqref{eq:eft-P}, and using the two-form and six-form flux relations in \eqref{eq:eft-W0}, we get 
\begin{equation}
 \mathcal P\big|_*=\frac{r_4\Delta}{12y}
 =\frac{e^{2D}\Delta}{12}
 =\frac1{2L_{3,E}},
 \qquad
 e^{K/2}W_0\big|_*=2\mathcal P\big|_*,
 \label{eq:eft-Pvac}
\end{equation}
while the K\"ahler covariant derivatives likewise satisfy
\begin{align}
 e^{K/2}D_{X^A}W_0\big|_*
 &=-\left.\frac{Q_A}{4yx_A^2}\right|_*
 =2\partial_{X^A}\mathcal P\big|_*,
 \\
  D_{T_a}W_0\big|_*&=0=\partial_{T_a}\mathcal P|_*,\\
 e^{K/2}D_YW_0\big|_*
 &=-\left.\frac{\mathcal P}{y}\right|_*
 =2\partial_Y\mathcal P\big|_*.
 \label{eq:eft-BPS}
\end{align}
Thus both conditions in \eqref{eq:eft-susy-conditions} are satisfied: the $r=2$ sector preserves supersymmetry and evaluating \eqref{eq:eft-V} on the vacuum also recovers the cosmological constant
\begin{equation}
 V\big|_*=-4\mathcal P\big|_*^2
 =-\frac{e^{4D}\Delta^2}{36}
 =-\frac1{L_{3,E}^2},
 \label{eq:eft-vacuumdata}
\end{equation}
in agreement with the dimensional reduction. The vacuum also has two physical RR axionic flat directions. Indeed, the potential depends on the four RR axions only through their sum $\Xi$, so three independent combinations can vary without changing the potential. One of these variations, $\delta\xi_A=Q_A\lambda$, is a gauge transformation because $\sum_AQ_A=0$, leaving two physical flat directions. The additional axion $\upsilon$ is independently removed by its own gauge symmetry.

To test the other sector, we evaluate the gravitino-shift
matrix on the same vacuum,
\begin{equation}
 A_1\big|_*
 =
 \begin{pmatrix}
 -3&0\\
 0&1
 \end{pmatrix}\mathcal P\big|_*.
 \label{eq:eft-A1vac}
\end{equation}
The first eigenvalue has magnitude $3/(2L_{3,E})$, rather than the value $1/(2L_{3,E})$ required for a preserved supersymmetry. The spin-$1/2$ shifts are also nonzero, since \eqref{eq:eft-BPS} gives
\begin{equation}
 (A_2)_I^{11}\big|_*
 =-\partial_I\mathcal P\big|_*,
 \qquad \text{but} \qquad
 \partial_Y\mathcal P\big|_*
 =-\left.\frac{\mathcal P}{2y}\right|_*\neq0.
\end{equation}
Therefore, the vacuum realizes $\mathcal N=2\to\mathcal N=1$ supersymmetry breaking in the gauged theory. 
The gravitino associated with the broken supersymmetry has mass parameter $|m_{3/2,\mathrm{broken}}|=3/(2L_{3,E})$, so supersymmetry breaking occurs at the anti-de Sitter scale. In the scale-separated regime of section~\ref{sec:scale-separation}, using the estimate for the Kaluza-Klein scale adopted there, we obtain
\begin{equation}
 \frac{|m_{3/2,\mathrm{broken}}|}{m_{\mathrm{KK}}}
 \sim\frac{r_{\max}}{L_{\mathrm{AdS}}}
 \longrightarrow0.
\end{equation}
Partial supersymmetry breaking therefore occurs parametrically below the Kaluza--Klein scale, within the lower-dimensional effective theory.

\section{Conclusions}
\label{sec:conclusions}

In this work, we have proposed new candidate string theory vacua with scale separation in which partial supersymmetry breaking occurs parametrically below the Kaluza--Klein scale. The construction gives three-dimensional anti-de Sitter solutions of massless type IIA supergravity with smeared O6-planes, whose effective action is reproduced by a three-dimensional $\mathcal N=2$ gauged supergravity with an $\mathcal N=1$ anti-de Sitter vacuum. Scale separation and partial supersymmetry breaking are therefore realized dynamically and simultaneously, allowing the latter to be studied within the lower-dimensional effective description.

The ten-dimensional solutions have an $\mathrm{SU}(3)$ structure and form families labeled by certain unbounded flux parameters. Along these families, we identified limits in which all internal radii become parametrically large in string units, the curvature becomes small and the anti-de Sitter radius grows faster than every internal radius. These limits include both weak and strong string coupling. In the weakly coupled regime, the solutions satisfy the usual bulk conditions for control of the supergravity approximation. At strong coupling, a suitable restriction on the flux scalings also makes the internal lengths large in eleven-dimensional Planck units, while keeping the Kaluza-Klein states of the M-theory circle above the anti-de Sitter scale. This provides a preliminary consistency check for a possible eleven-dimensional uplift. Recent discussions on M-theory superpotentials and the challenges of such compactifications can be found for example in \cite{Sethi:2017phn,Ashmore:2026vgf}. 

Our work suggests several directions for further investigation. A central open problem is to understand the sources beyond the smeared approximation and account for their full backreaction. We believe that our solutions may offer an advantage in this respect because, as for their four-dimensional counterparts \cite{Cribiori:2021djm}, they may allow this problem to be addressed geometrically in eleven dimensions. It would therefore be important to investigate their possible uplift to novel candidate M-theory vacua with scale separation. Another approximation concerns the geometric estimate of the Kaluza--Klein scale, which should ultimately be supplemented by a more precise spectral analysis.
More broadly, our results suggest that scale separation may favor minimal supersymmetry through a nontrivial interplay between partial supersymmetry breaking and restrictions imposed by the geometry and sources. It would be important to establish whether this apparent correlation between scale separation and minimal, or even no, supersymmetry is a lamppost effect or a consequence of a more general principle of quantum gravity.

\paragraph{Acknowledgments}
The work of N.C.~is supported by the Research Foundation Flanders (FWO grant 1259125N).
We acknowledge the use of large language models to assist with editing and checking calculations in this manuscript. We take full responsibility for the content, accuracy, and conclusions of the work.

\appendix

\section{O6-plane charge and tadpole cancellation}
\label{app:O6count}

In this appendix, we determine the O6-plane charge $Q_{\rm O6}$ entering the tadpole condition $Q_{\rm O6}=-m_0h$ of section~\ref{sec:fluxes}. This requires counting the O6-planes, namely the components of the fixed loci of the orientifold involution, which on the twisted torus depends on the parity
of the metric flux $h$.

We work on the covering space $\tilde T^6$ of the orbifold. The circle direction $\eta^4$ is invariant under all involutions and is wrapped by all O6-planes, so it plays no role in what follows. On $\tilde T^6/\mathbb{Z}_2\times\mathbb{Z}_2$, the four involutions induce the same map, since they differ by orbifold elements, and we can thus consider $\sigma_\alpha$ only. A point of the orbifold is fixed by $\sigma_\alpha$ if $\sigma_\alpha$ maps it to itself up to an orbifold transformation, so that on $\tilde T^6$ the O6-planes correspond to the fixed loci of all four involutions $\sigma_\alpha$, $\sigma_{\alpha\beta}$, $\sigma_{\alpha\gamma}$ and $\sigma_{\alpha\beta\gamma}$. 
Since all fields and sources are invariant under the orbifold, the Bianchi identity $dF_2=j_{\rm O6}$ can be equivalently imposed on $\tilde T^6$.

We introduce coordinates $x^i$ on the universal cover $\mathbb{R}^6$ and represent the left-invariant one-forms as
\begin{equation}
    \eta^1=dx^1+h\left(x^2dx^7+x^3dx^6\right),\qquad
    \eta^5=dx^5-h\left(x^6dx^7+x^2dx^3\right),
    \label{eq:app-eta}
\end{equation}
and $\eta^i=dx^i$ otherwise, which reproduce the Maurer--Cartan equations \eqref{eq:MC}. The twisted torus $\tilde T^6$ is obtained by identifying $x\sim n\cdot x$ for all $n\in\mathbb{Z}^6$, where
\begin{equation}
\begin{aligned}
    (n\cdot x)^1&=x^1+n^1-h\left(n^2x^7+n^3x^6\right),\\
    (n\cdot x)^5&=x^5+n^5+h\left(n^6x^7+n^2x^3\right),
\end{aligned}
\label{eq:app-product}
\end{equation}
with $(n\cdot x)^i=x^i+n^i$ for the remaining coordinates.\footnote{These identifications are fixed by requiring the forms \eqref{eq:app-eta} to be invariant. Shifting the base coordinates as $x^i\to x^i+n^i$, $i=2,3,6,7$, leaves $\eta^i=dx^i$ invariant. Writing $x^1\to x^1+n^1+f(x)$, one finds $\eta^1\to\eta^1+df+h\left(n^2dx^7+n^3dx^6\right)$, so that invariance requires $f=-h\left(n^2x^7+n^3x^6\right)$ up to a constant. Similarly, invariance of $\eta^5$ fixes the shift of $x^5$.} This identification is consistent precisely when $h\in\mathbb{Z}$, since only then the composition of two identifications with $n,n'\in\mathbb{Z}^6$ is again an identification with integer components. Each involution acts on the coordinates by flipping the sign of the transverse directions of table~\ref{tab:O6}, for instance $\sigma_\alpha:(x^1,x^6,x^7)\mapsto-(x^1,x^6,x^7)$. One can check that $\sigma(n\cdot x)=n'\cdot\sigma(x)$, where $n'\in\mathbb{Z}^6$ is obtained from $n$ by flipping the signs of the same components as $\sigma$, so that $\sigma$ is well defined on $\tilde T^6$.

A point of $\tilde T^6$ is fixed by $\sigma$ if $\sigma(x)=n\cdot x$ for some $n\in\mathbb{Z}^6$. Let us consider $\sigma_\alpha$ in detail. Component by component, the condition gives
\begin{equation}
\begin{aligned}
    &x^1=-\frac{1}{2}n^1\,,\qquad x^6=-\frac{1}{2}n^6\,,\qquad
    x^7=-\frac{1}{2}n^7\,,\\
    &n^5=\frac{h}{2}\,n^6n^7\,,\,\,\quad n^2=0\,,\qquad  \qquad \,\,n^3=0\,,
\end{aligned}
\label{eq:sigmaafixedpt}
\end{equation}
while $x^2$, $x^3$ and $x^5$ are unconstrained. The fixed loci are thus
located at the eight transverse positions $(x^1,x^6,x^7)\in\{0,\frac12\}^3$ and wrap the directions $(\eta^2,\eta^3,\eta^5)$. Notice that, since $\eta^5$ restricts to $dx^5-h\,x^2dx^3$, they are three-dimensional Heisenberg nilmanifolds, and each transverse position contributes a unit of coordinate volume, $\int\eta^{235}=1$. Since $n^5\in\mathbb{Z}$, the condition \eqref{eq:sigmaafixedpt} requires $hn^6n^7$ to be even. For $h$ even, this is always satisfied and all eight transverse positions give fixed loci. For $h$ odd, the two positions with $x^6=x^7=\tfrac12$ are to be excluded since $hn^6n^7$ is odd (indeed, from \eqref{eq:sigmaafixedpt} we see that $n^6$ and $n^7$ must be odd in this case). The total wrapped volume of the fixed loci of $\sigma_\alpha$ is thus $8$ for $h$ even and $6$ for $h$ odd. Since O6-planes carry a fixed charge per unit volume, this is the quantity that determines the smeared charge.

The other three involutions can be analysed in the same way. Each of them
flips one fiber direction and two base directions, and the fixed-point
condition again constrain $h/2$ times the
product of the two components of $n$ along the flipped base directions to be integer,
namely
\begin{equation}
    \sigma_{\alpha\beta}:\ n^1=-\frac{h}{2}\,n^2n^7\,,\qquad
    \sigma_{\alpha\gamma}:\ n^1=-\frac{h}{2}\,n^3n^6\,,\qquad
    \sigma_{\alpha\beta\gamma}:\ n^5=\frac{h}{2}\,n^2n^3\,.
\end{equation}
As for $\sigma_\alpha$, all eight transverse positions are allowed for $h$
even, while two of them are excluded for $h$ odd. Therefore, for each
involution the total wrapped coordinate volume is
\begin{equation}
    n_{\rm O6}=\begin{cases}
    8\,, & h\ \text{even}\,,\\
    6\,, & h\ \text{odd}\,.
    \end{cases}
\end{equation}
The case of even $h$ reproduces the counting on an ordinary torus, while for odd $h$ two of the eight O6-planes associated with each involution are missing. A similar subtlety was pointed out in \cite{Frey:2002hf} for odd fluxes on toroidal orientifolds, where it is resolved by additional discrete flux at the fixed loci, in line with the classification of exotic O-planes in \cite{Hanany:2000fq}. To avoid this subtlety, we can take $h$ to be even.

In units $2\pi\sqrt{\alpha'}=1$, a D6-brane carries unit charge and an O6$^-$-plane carries charge $-2$ on the covering space. Hence, upon smearing, $Q_{\rm O6}=-2\,n_{\rm O6}$, and the tadpole condition $Q_{\rm O6}=-m_0h$ becomes
\begin{equation}
    m_0\,h=16\,,
\end{equation}
for $h$ even, or $m_0h=12$ for $h$ odd. This admits integer solutions with $m_0,h>0$, as in the vacua of section~\ref{sec:3dvacuum}, for instance $(h,m_0)=(2,8)$ or $(4,4)$. Since $m_0$ and $h$ enter the vacua only through order-one factors, their precise values do not affect any of the parametric statements of the main text. Including $n_{\rm D6}$ D6-branes parallel to the O6-planes would replace $Q_{\rm O6}$ with the net charge $-2\,n_{\rm O6}+n_{\rm D6}$, thus enlarging the set of solutions to the tadpole condition, at the cost of introducing open-string moduli. In this work, we do not include D6-branes.

\section{Direct dimensional reduction}
\label{app:dimensional-reduction}

We derive the three-dimensional bosonic effective action of the retained left-invariant sector in the smeared approximation, including the axions and the circle vector from the outset. After specifying the reduction ansatz and the resulting three-dimensional action, we derive its scalar kinetic terms, vector kinetic term and scalar potential separately. We finally give the equivalent first-order Chern--Simons formulation used in the main text.

\subsection{Conventions and reduction ansatz}
\label{app:dr-conventions}

The metric, radii $r_i$ and the dressed metric fluxes $a_i$ introduced in section~\ref{sec:themodel} are defined in the ten-dimensional string frame. We keep this notation and explicitly label their Einstein-frame counterparts as
\begin{equation}
 g^{(10,E)}_{MN}=e^{-\phi/2}g^{(10,S)}_{MN},\qquad
 r_{i,E}=e^{-\phi/4}r_i,\qquad
 a_{i,E}=e^{\phi/4}a_i.
 \label{eq:dr-frame-dictionary}
\end{equation}
Here, the dilaton $\phi$ and the radii are independent of the internal coordinates, but may depend on the three external coordinates. We also define the volume functions
\begin{equation}
 \mathcal V_{6,S}=r_1r_2r_3r_5r_6r_7=t_1t_2t_3,\qquad
 \mathcal V_{7,S}=r_4\mathcal V_{6,S},
 \label{eq:V6sV7s}
\end{equation}
so that
\begin{equation}
 \mathcal V_{6,E}=e^{-3\phi/2}\mathcal V_{6,S},\qquad
 \mathcal V_{7,E}=e^{-7\phi/4}\mathcal V_{7,S}.
 \label{eq:dr-volumes}
\end{equation}

The only orbifold-invariant one-form is $\eta^4=dy^4$, which is even under the orientifold. The reduction of the ten-dimensional metric therefore gives rise to a vector $\mathcal A$ associated to the circle direction. We include it by defining
\begin{equation}
 \vartheta=\eta^4+\mathcal A = dy^4 + \mathcal A,\qquad \mathcal F_2=d\mathcal A
 \label{eq:dr-circle-connection}
\end{equation}
and then taking the Einstein-frame reduction ansatz 
\begin{equation}
 ds^2_{10,E}=\mathcal V_{7,E}^{-2}ds^2_{3,E}
 +r_{4,E}^2\vartheta^2+\sum_{i\neq4}r_{i,E}^2(\eta^i)^2.
 \label{eq:dr-metric}
\end{equation}
\label{eq:dr-vector-metric}
Equivalently, in string-frame
\begin{equation}
 ds^2_{10,S}=e^{4D}ds^2_{3,E}
 +r_4^2\vartheta^2+\sum_{i\neq4}r_i^2(\eta^i)^2,
 \qquad e^{-2D}=e^{-2\phi}\mathcal V_{7,S},
 \label{eq:dr-string-metric}
\end{equation}
where $D=\phi-\tfrac12\log\mathcal V_{7,S}$ is the three-dimensional
dilaton.  On the anti-de Sitter vacuum, $\mathcal A=0$ and $D=D_0$ is constant, so
\begin{equation}
 ds^2_{3,E}=e^{-4D_0}ds^2_{\mathrm{AdS}_3},\qquad
 L_{3,E}=e^{-2D_0}L_{\rm AdS}.
 \label{eq:dr-AdS-radius}
\end{equation}

In the standard, non-democratic formulation, the ten-dimensional bulk terms relevant to us are \cite{Bergshoeff:2001pv,VanRiet:2023pnx}
\begin{equation}
\begin{split}
 S_{10}=\frac{1}{2\kappa_{10}^2}\int d^{10}x\sqrt{-g_{10,E}}\,
 \Big[R_{10,E}-\frac12(\partial\phi)^2
       -\frac12e^{-\phi}|H_3|_{E}^2\\
 {}-\frac12 e^{3\phi/2}|F_2|_{E}^2
       -\frac12 e^{\phi/2}|F_4^{(10)}|_{E}^2\Big]
 +S_{\mathrm{O6}}+S_{\mathrm{top}}.
\end{split}
 \label{eq:dr-action}
\end{equation}
We set $|F_p|^2=F_{M_1\ldots M_p}F^{M_1\ldots M_p}/p!$, and frame labels on norms specify which ten-dimensional metric is used in the contraction; external indices are contracted with $g_{3,E}$ unless stated otherwise.  $F_4^{(10)}$ includes both internal and electric components; $S_{\rm top}$ denotes the bulk topological terms.
Recall that in our conventions $2\pi\ell_s=1$, with $\ell_s=\sqrt{\alpha'}$ and
$(2\kappa_{10}^2)^{-1}=2\pi/(2\pi\ell_s)^8$.
The DBI part of the source action is
\begin{equation}
 S_{\mathrm{O6}}=-\sum_I T_I
 \int_{{\rm AdS}_3\times\Sigma_I}d^7\xi\,
 e^{3\phi/4}\sqrt{-\det P_I[g_{10,E}]},
 \qquad
 \Sigma_I:\quad2345,\;1346,\;1247,\;4567,
 \label{eq:dr-source-action}
\end{equation}
where $P_I[g]$ is the induced metric and $T_I<0$ is the total tension for the O6-planes wrapping the cycle $\Sigma_I$. In our conventions, for these BPS sources  $2\kappa_{10}^2T_I=Q_I$ and then the Bianchi identity fixes $Q_I=Q_{\mathrm{O6}}=-hm_0$ for each of the four cycle types.

\paragraph{Invariant forms and field expansions.}
We denote the two-forms underlying
\eqref{Kahler_expansion} by
\begin{equation}
 \omega_1=-\eta^{15},\qquad
 \omega_2=-\eta^{26},\qquad
 \omega_3=\eta^{37}
\end{equation}
which are orientifold odd, while the three-forms 
\begin{equation}
 \alpha_0=\eta^{127},\qquad
 \alpha_1=\eta^{136},\qquad
 \alpha_2=-\eta^{235},\qquad
 \alpha_3=-\eta^{567}
 \label{eq:alpha-basis}
\end{equation}
are orientifold even; the circle one-form $\eta^4=dy^4$
is also even. Let us define
\begin{equation}
 \beta_0=\eta^{257}+\eta^{356}-\eta^{167}-\eta^{123},
 \qquad
 \widehat{\mathrm{vol}}_6
 =\omega_1\wedge\omega_2\wedge\omega_3
 =-\eta^{123567},
\end{equation}
where $\widehat{\mathrm{vol}}_6$ is independent of the moduli and satisfies $\mathrm{vol}_6=\mathcal V_{6,S}\widehat{\mathrm{vol}}_6$.
Below we will need the following properties ($A=0,1,2,3$, $a=1,2,3$)
\begin{align}
 d\omega_1&=h\beta_0,
 &d\omega_2&=d\omega_3=0,
 &d\alpha_A&=h\omega_2\wedge\omega_3,\nonumber\\
 \omega_a\wedge\alpha_A&=0,
 &\alpha_A\wedge\beta_0&=\widehat{\mathrm{vol}}_6,
 &\omega_a\wedge\omega_a&=0=\omega_a\wedge\beta_0.
 \label{eq:dr-basis-identities}
\end{align}

The orientifold parities of $B_2,C_1,C_3$ are respectively
odd, odd and even. Accordingly, we expand the fields as
\begin{align}
 B_2&=-\sum_{a=1,2,3} b_a(x)\omega_a,\nonumber\\
 C_3
 &=\sum_{A=0,1,2,3}\xi_A(x)\alpha_A+y^4\mathcal Q_3
   +\mathcal D_2(x)\wedge\vartheta+\mathcal C_3(x),
 \label{eq:dr-axion-conventions}
\end{align}
where $\mathcal D_2$ and $\mathcal C_3$ are external
two- and three-forms, and
\begin{equation}
 \mathcal Q_3=\sum_AQ_A\alpha_A,
 \qquad
 (Q_0,Q_1,Q_2,Q_3)=(f_3,f_1,f_2,f_4).
\end{equation}
Here, $C_3$ is the potential entering $F_4^{(10)}=dC_3+B_2\wedge F_2$ locally.
The term $y^4\mathcal Q_3$ changes by $\mathcal Q_3$ under one loop around the circle, $y^4\to y^4+1$. Since the Bianchi identity \eqref{eq:dF4tad}, now rewritten as $\sum_AQ_A=0$, implies $d\mathcal Q_3=0$, this change leaves the field strength invariant. 
The coefficients $b_a$ and $\xi_A$ give three NS and four RR axions, respectively. 
The external two-form $\mathcal D_2$ gives the electric flux $F_4^{\mathrm{el}}=d_3\mathcal D_2\wedge\vartheta$. It has no local propagating degrees of freedom, but integrating it out produces a contribution to the scalar potential. The purely external three-form $\mathcal C_3$ has no field-strength kinetic term in three dimensions.
Since there are no invariant odd one-forms or invariant even two-forms, $B_2$ and $C_3$ cannot supply additional three-dimensional vectors, while $C_1$ has no surviving scalar or vector fluctuations. We nevertheless include its nonzero background two-form flux as
\begin{equation}
 F_2=-m_0\omega_1-N\omega_2-R_F\omega_3.
\end{equation}

The internal RR four-form flux $F_4^{\mathrm{bg}}=\eta^4\wedge\mathcal Q_3$ (with $\eta^4 = dy^4$) gauges the axionic shifts through the Kaluza--Klein vector $\mathcal A$. Indeed, under the local circle transformation $y^4\mapsto y^4-\lambda(x)$, invariance of $\vartheta=\eta^4+\mathcal A$ and of the combination $\sum_A\xi_A\alpha_A+y^4\mathcal Q_3$ requires $ \delta\mathcal A=d\lambda$ and $ \delta\xi_A=Q_A\lambda$.
The corresponding gauge-invariant derivatives are therefore
\begin{equation}
 D\xi_A=d\xi_A-Q_A\mathcal A.
 \label{eq:dr-axion-gauging}
\end{equation}
Moreover, $\Xi=\sum_A\xi_A$ is gauge invariant because $\sum_AQ_A=0$.
From the expansions of the gauge potentials and using \eqref{eq:dr-basis-identities}, we obtain 
\begin{align}
 H_3&=-\sum_a db_a\wedge\omega_a-hb_1\beta_0,
 \nonumber\\
 F_4^{(10)}
 &= \vartheta\wedge\mathcal Q_3 
   +\sum_A D\xi_A\wedge\alpha_A
   +\mathcal G_4+\mathcal H_3\wedge\vartheta,
 \label{eq:dr-full-field-strengths}
\end{align}
where
\begin{equation}
\mathcal G_4
 =h\Xi\omega_2\wedge\omega_3+B_2\wedge F_2,
 \qquad
 \mathcal H_3=d\mathcal D_2.
\end{equation}
Here, $\mathcal G_4$ is the internal axion-dependent four-form, while $\mathcal H_3\wedge\vartheta$ is the electric component; the terms $d_3\mathcal C_3$ and $\mathcal D_2\wedge d\mathcal A$ vanish identically, since they have four external legs.
The resulting field strengths satisfy the Bianchi identity $dF_4^{(10)}=H_3\wedge F_2$. In particular, writing $\mathcal F_2=d\mathcal A$, the terms involving its field strength cancel:
\begin{equation}
 d\vartheta\wedge\mathcal Q_3
 +\sum_A d(D\xi_A)\wedge\alpha_A
 =\mathcal F_2\wedge\mathcal Q_3
 -\sum_A Q_A\mathcal F_2\wedge\alpha_A=0.
\end{equation}
The remaining terms give $H_3\wedge F_2$; the possible source contribution $B_2\wedge dF_2$ vanishes because $dF_2=-hm_0\beta_0$ and $\omega_a\wedge\beta_0=0$.

The geometric scalar fields are defined by
$J=\sum_a t_a\omega_a$ and
\begin{equation}
\begin{aligned}
 t_1&=r_1r_5,\qquad
 t_2=r_2r_6,\qquad
 t_3=r_3r_7,\\
 (x_0,x_1,x_2,x_3)
 &=e^{-\phi}(r_1r_2r_7,r_1r_3r_6,
             r_2r_3r_5,r_5r_6r_7),\\
 y&=e^{-2\phi}r_4^2\mathcal V_{6,S}.
\end{aligned}
 \label{eq:dr-scalar-dictionary}
\end{equation}
They combine with the axions into the complex scalars $X^A=x_A+i\xi_A$ and  $T_a=t_a+i b_a$. The scalar $\upsilon$ completing $Y=y+i\upsilon$ will be introduced below when rewriting the vector sector in an equivalent first-order formulation. We will denote $g_s=e^\phi$.

\paragraph{Three-dimensional action.}
For clarity, we first state the three-dimensional action that we will derive in the following subsections.
In the formulation with a propagating Kaluza--Klein vector, and after eliminating the external two-form, it takes the form
\begin{equation}
 S_{3D}=\frac1{\kappa_3^2}\int d^3x\sqrt{-g_{3,E}}
 \left[\frac12R_{3,E}-\mathcal L_{\mathrm{scal}}
 -\frac{y^2}{8}\mathcal F_{\mu\nu}\mathcal F^{\mu\nu}
 -V_{\mathrm{full}}\right],
 \label{eq:dr-3d-action}
\end{equation}
with
\begin{equation}
 \mathcal L_{\mathrm{scal}}=\mathcal{L}_{\rm sax}+ \mathcal{L}_{\rm ax}=
 \sum_A\frac{1}{4x_A^2}((\partial x_A)^2+(D\xi_A)^2)
 +\sum_a\frac{1}{4t_a^2}((\partial t_a)^2+(\partial b_a)^2)
 +\frac{(\partial y)^2}{4y^2}.
 \label{eq:dr-Maxwell-description}
\end{equation}
Here, $(D\xi_A)^2=D_\mu\xi_A D^\mu\xi_A$, and similarly for the other kinetic terms. We derive the scalar and vector kinetic terms below, followed by the full potential \eqref{eq:dr-full-axion-potential}, including the contribution from eliminating the external two-form.
Section~\ref{app:dr-CS} gives an equivalent first-order description in which the Maxwell term is replaced by an auxiliary-field formulation containing a BF coupling.

\subsection{Kinetic terms}
We first work out the kinetic terms of scalars and vectors. 

\subsubsection{Scalar kinetic terms}
\label{app:dr-scalar-kinetic}

To derive the kinetic terms of the real scalar moduli, we set the Kaluza--Klein vector to zero and write the metric as
\begin{equation}
 ds^2_{10,E}=d\bar s_3^2
 +\sum_{i=1}^7e^{2\rho_i(x)}(\eta^i)^2,
 \qquad
 \rho_i=\log r_{i,E},
 \qquad
 s=\sum_i\rho_i=\log\mathcal V_{7,E}.
\end{equation}
The external derivative terms in the Ricci scalar have the same form as those of a multiply warped product with one-dimensional fibres ($s_i=1$) and warping functions $b_i=e^{\rho_i}$, in the notation of  \cite[Proposition~2.6, Item (1)]{Dobarro:2004iz}.
The internal derivatives of the left-invariant coframe contribute the intrinsic curvature $R_{7,E}$, so that
\begin{equation}
 R_{10,E} =\bar R_3+R_{7,E}  -2\bar\Box s-(\bar\partial s)^2 -\sum_i(\bar\partial\rho_i)^2.
\end{equation}
Passing to the three-dimensional Einstein frame through $\bar g_{\mu\nu}=e^{-2s}g_{\mu\nu,E}$ gives
\begin{equation}
 R_{10,E}
 =R_{7,E}
 +e^{2s}\left[
 R_{3,E}+2\Box s-(\partial s)^2
 -\sum_i(\partial\rho_i)^2
 \right],
\end{equation}
where unbarred derivatives are contracted with $g_{3,E}$.
After integration over the internal space, the measure contributes $e^{-2s}\sqrt{-g_{3,E}}$, making the $\Box s$ term a total derivative. Discarding this term and including the ten-dimensional dilaton kinetic term we obtain, with the normalization of \eqref{eq:dr-3d-action}, a contribution $\mathcal{L}_{\rm sax}$ to $\mathcal{L}_{\rm scal}$ which reads 
\begin{equation}
 \mathcal L_{\mathrm{sax}}
 =\frac12\sum_i(\partial\rho_i)^2
 +\frac12(\partial s)^2
 +\frac14(\partial\phi)^2.
\end{equation}
Using $\rho_i=\log r_i-\phi/4$ and $D=\phi-\frac12\log\mathcal V_{7,S}$, this becomes
\begin{equation}
 \mathcal L_{\mathrm{sax}}
 =\frac12\sum_i(\partial\log r_i)^2
 +2(\partial D)^2.
 \label{eq:dr-kinetic}
\end{equation}
Finally, expressing the radii and dilaton in terms of the scalar coordinates \eqref{eq:dr-scalar-dictionary} gives 
\begin{equation}
 \mathcal L_{\mathrm{sax}}
 =\frac14\sum_A(\partial\log x_A)^2
 +\frac14\sum_a(\partial\log t_a)^2
 +\frac14(\partial\log y)^2.
\end{equation}

The components $D\xi_A\wedge\alpha_A$ of $F_4^{(10)}$ and $-db_a\wedge\omega_a$ of $H_3$, each with one external index, give the axion kinetic terms. For the diagonal internal metric, the squared norm of each basis form is the inverse squared product of the corresponding radii; for example, $\alpha_0=\eta^{127}$ and $x_0=e^{-\phi}r_1r_2r_7$ imply $|\alpha_0|_{6,S}^2=e^{-2\phi}/x_0^2$.
Moreover, the frame relation $g_{7,E}=e^{-\phi/2}g_{7,S}$ gives $|\alpha_p|_E^2=e^{p\phi/2}|\alpha_p|_S^2$ for an internal $p$-form. Hence, we have
\begin{equation}
 e^{\phi/2}|\alpha_A|_{6,E}^2
 =e^{2\phi}|\alpha_A|_{6,S}^2
 =\frac1{x_A^2},
 \qquad
 e^{-\phi}|\omega_a|_{6,E}^2
 =|\omega_a|_{6,S}^2
 =\frac1{t_a^2}.
\end{equation}
The single external index contributes a factor $\mathcal V_{7,E}^{2}$ from the inverse metric, cancelling $\mathcal V_{7,E}^{-2}$ from the reduced measure. Thus, with the action normalization of \eqref{eq:dr-3d-action}, the ten-dimensional RR and NS kinetic terms give
\begin{equation}
 \mathcal L_{\mathrm{ax}}
 =\sum_A\frac{(D\xi_A)^2}{4x_A^2}
  +\sum_a\frac{(\partial b_a)^2}{4t_a^2}.
 \label{eq:dr-axion-kinetic}
\end{equation}
The basis forms are mutually orthogonal with respect to the diagonal internal metric, so no cross terms occur between distinct axions in these kinetic terms. Adding the saxion contribution \eqref{eq:dr-kinetic} gives \eqref{eq:dr-Maxwell-description}.

\subsubsection{Vector kinetic term}
\label{app:vector-reduction}

The Kaluza--Klein vector enters the Einstein--Hilbert term through the curvature of the circle fibration\cite[eq.~(2.5)]{Maharana:1992my}
\begin{equation}
 R_{10,E}\supset
 -\frac14r_{4,E}^2\mathcal V_{7,E}^{4}
 \mathcal F_{\mu\nu}\mathcal F^{\mu\nu},
\end{equation}
where indices are raised with $g_{3,E}$. The factor $\mathcal V_{7,E}^{4}$ comes from the two inverse external metrics needed to contract $\mathcal F_2 = d \mathcal{A}$. The reduced measure contributes $\mathcal V_{7,E}^{-2}$,so, with the normalization of \eqref{eq:dr-3d-action}, we obtain
\begin{equation}
 S_{\mathrm{KK}}
 =-\frac1{\kappa_3^2}\int d^3x\sqrt{-g_{3,E}}\,
 \frac{\mathcal V_{7,E}^{2}r_{4,E}^{2}}8
 \mathcal F_{\mu\nu}\mathcal F^{\mu\nu}
 =-\frac1{\kappa_3^2}\int d^3x\sqrt{-g_{3,E}}\,
 \frac{y^2}{8}\mathcal F_{\mu\nu}\mathcal F^{\mu\nu},
 \label{eq:dr-KK-Maxwell}
\end{equation}
where in the last equality we used $\mathcal V_{7,E}^{2}r_{4,E}^{2} =e^{-4\phi}r_4^4\mathcal V_{6,S}^{2}=y^2$.
This reproduces the Maxwell term in \eqref{eq:dr-3d-action} for the circle connection $\vartheta=\eta^4+\mathcal A$.

The vector acquires a mass through its coupling to the RR axions, $D\xi_A=d\xi_A-Q_A\mathcal A$: one axionic combination supplies its longitudinal polarization through the Stückelberg mechanism.\footnote{All O6 sources wrap the circle. Since the determinant of their induced metric is independent of $\mathcal A$, their DBI action gives no additional vector terms.}
The canonically normalized vector on the vacuum is $\mathcal A^{\rm can}=y\mathcal A/\sqrt2$. Comparing the quadratic term from the axion kinetic terms with $-\frac12m_{\mathcal A}^2(\mathcal A^{\rm can})^2$ gives
\begin{equation}
 m_{\mathcal A}^2
 =\frac1{y^2}\sum_A\frac{Q_A^2}{x_A^2},
\end{equation}
On the general solution, the flux relations \eqref{eq:flux-radius-relations}, together with the Bianchi identity \eqref{eq:dF4tad}, imply $\sum_AQ_A^2/x_A^2=r_4^2\Delta^2/9$. Using \eqref{eq:dr-scalar-dictionary}, \eqref{eq:dr-string-metric} and \eqref{eq:dr-AdS-radius}, we therefore obtain
\begin{equation}
 m_{\mathcal A}^2
 =\frac{r_4^2\Delta^2}{9y^2}
 =e^{4D}\frac{\Delta^2}{9}
 =\frac4{L_{3,E}^2}.
\end{equation}
Thus the retained vector has a mass of order the anti-de Sitter scale.

\subsection{Scalar potential}
\label{app:dr-potential}

The scalar potential receives contributions from the internal curvature, the internal NS and RR field strengths, the external two-form $\mathcal D_2$ and the O6 tensions. The field strength $\mathcal H_3=d_3\mathcal D_2$ supplies the electric component $F_4^{\mathrm{el}}=\mathcal H_3\wedge\vartheta$ of the ten-dimensional RR four-form. We derive these contributions in the normalization of \eqref{eq:dr-3d-action}. As in the preceding subsections, the reduced measure supplies a factor $\mathcal V_{7,E}^{-2}$.

\paragraph{Internal curvature}

To compute the internal Ricci scalar, we express the Maurer--Cartan
equations \eqref{eq:MC} in the orthonormal coframe,
$de_E^i=-\frac12f^i{}_{jk,E}e_E^j\wedge e_E^k$.
Since the Lie algebra is nilpotent, it is unimodular and its
Killing form vanishes. The curvature contribution in
\cite[eq.~(5.14)]{Hull:2005hk} therefore reduces to
\begin{equation}
 R_{7,E}=-\frac14\sum_{i,j,k}(f^i{}_{jk,E})^2
        =-\frac12\sum_{i=1}^4a_{i,E}^2.
\end{equation}
The second equality follows because the four independent structure constants have magnitudes $|a_{i,E}|$, and each occurs twice in the sum over the antisymmetric indices $j,k$. Since the action contains $R_{3,E}/2-V$, the curvature contribution to the potential is
\begin{equation}
 V_R=-\frac{R_{7,E}}{2\mathcal V_{7,E}^2}
    =\frac1{4\mathcal V_{7,E}^2}\sum_{i=1}^4a_{i,E}^2
    =\frac{e^{4D}}4\sum_{i=1}^4a_i^2,
 \label{eq:dr-curvature}
\end{equation}
where the last equality uses \eqref{eq:dr-frame-dictionary} and \eqref{eq:dr-string-metric}.

\paragraph{Internal fluxes}

The internal two-form flux gives
\begin{equation}
 V_{F_2}
 =\frac{e^{3\phi/2}}{4\mathcal V_{7,E}^2}|F_2|_{7,E}^2
 =\frac{e^{4D}g_s^2}{4}
 \left(\frac{m_0^2}{t_1^2}+\frac{N^2}{t_2^2}
       +\frac{R_F^2}{t_3^2}\right),
 \label{eq:dr-F2E}
\end{equation}
where we used \eqref{RR2_expansion}, \eqref{eq:dr-string-metric} and \eqref{eq:dr-scalar-dictionary}. Likewise, the background four-form $F_4^{\mathrm{bg}}=\eta^4\wedge\mathcal Q_3$ in \eqref{eq:dr-full-field-strengths} contributes
\begin{equation}
 V_{F_4}
 =\frac{e^{\phi/2}}{4\mathcal V_{7,E}^2}
       |F_4^{\mathrm{bg}}|_{7,E}^2
 =\frac{e^{4D}g_s^2}{4}|F_4^{\mathrm{bg}}|_{7,S}^2
 =\frac1{4y^2}\sum_A\frac{Q_A^2}{x_A^2},
 \label{eq:dr-F4E}
\end{equation}
where the last equality follows from \eqref{eq:dr-string-metric} and \eqref{eq:dr-scalar-dictionary}. The different basis components are orthogonal, so their squared coefficients add without cross terms.

For nonzero axions, the internal field strengths in
\eqref{eq:dr-full-field-strengths} give additional contributions
to the scalar potential. The internal NS three-form
$H_{3,\mathrm{int}}=-hb_1\beta_0$ has string-frame norm
\begin{equation}
 |H_{3,\mathrm{int}}|_{7,S}^2
 =h^2b_1^2\left(
 \frac1{r_2^2r_5^2r_7^2}
 +\frac1{r_3^2r_5^2r_6^2}
 +\frac1{r_1^2r_6^2r_7^2}
 +\frac1{r_1^2r_2^2r_3^2}\right).
\end{equation}
Likewise, expanding the axion-dependent four-form
$\mathcal G_4=h\Xi\omega_2\wedge\omega_3+B_2\wedge F_2$,
with $\Xi=\sum_A\xi_A$, gives
\begin{equation}
 \mathcal G_4
 =\mathcal R_{23}\omega_2\wedge\omega_3
 +\mathcal R_{13}\omega_1\wedge\omega_3
 +\mathcal R_{12}\omega_1\wedge\omega_2,
\end{equation}
where
\begin{equation}
 \mathcal R_{23}=h\Xi+Nb_3+R_Fb_2,\qquad
 \mathcal R_{13}=m_0b_3+R_Fb_1,\qquad
 \mathcal R_{12}=m_0b_2+Nb_1.
 \label{eq:dr-axion-polynomials}
\end{equation}
Using \eqref{eq:dr-scalar-dictionary}, its string-frame norm is
\begin{equation}
 |\mathcal G_4|_{7,S}^2
 =\frac{\mathcal R_{23}^2}{t_2^2t_3^2}
 +\frac{\mathcal R_{13}^2}{t_1^2t_3^2}
 +\frac{\mathcal R_{12}^2}{t_1^2t_2^2}.
\end{equation}
The corresponding contribution to the potential follows from
\eqref{eq:dr-action} and \eqref{eq:dr-string-metric}:
\begin{equation}
 V_{H_3}+V_{\mathcal G_4}
 =\frac{e^{4D}}4\left(
 |H_{3,\mathrm{int}}|_{7,S}^2
 +g_s^2|\mathcal G_4|_{7,S}^2\right).
\end{equation}
$F_4^{\mathrm{bg}}$ and $\mathcal G_4$ are orthogonal with respect to the diagonal internal metric and contribute no cross term to the potential.

\paragraph{Electric flux}

In the non-democratic action \eqref{eq:dr-action}, the internal
six-form flux is described by the electric component of
$F_4^{(10)}$. In our conventions, the corresponding duality
relation is
\begin{equation}
 (F_6)_{\mathrm{int}}
 =e^{\phi/2}*_{10,E}F_4^{\mathrm{el}},
 \label{eq:dr-duality}
\end{equation}
where we use the orientation $\mathrm{vol}_{10,E} =\mathcal V_{7,E}^{-3}\mathrm{vol}_{3,E} \wedge\mathrm{vol}_{7,E}$, with $\mathrm{vol}_{7,E}=e_E^4\wedge\mathrm{vol}_{6,E}$ and $\mathrm{vol}_{6,E}=-e_E^{123567}$; the factor $\mathcal V_{7,E}^{-3}$ comes from the external metric rescaling in \eqref{eq:dr-metric}.
In mostly-plus signature, this gives $*_{10,E}(\mathrm{vol}_{3,E}\wedge e_E^4) =-\mathcal V_{7,E}^{3}\mathrm{vol}_{6,E}$.
Writing the external field strength as $\mathcal H_3=d\mathcal D_2 =\mathfrak f\,\mathrm{vol}_{3,E}$, we have $F_4^{\mathrm{el}}=\mathcal H_3\wedge\vartheta$.
In the metric ansatz \eqref{eq:dr-metric}, the circle component of the ten-dimensional orthonormal frame is $e_E^4=r_{4,E}\vartheta$, reducing to $r_{4,E}\eta^4$ on the vacuum where $\mathcal A=0$.
Thus, we find
\begin{equation}
 F_4^{\mathrm{el}}
 =\frac{\mathfrak f}{r_{4,E}}\,
   \mathrm{vol}_{3,E}\wedge e_E^4,
 \qquad
 |F_4^{\mathrm{el}}|_{10,E}^2
 =-\mathfrak f^2\frac{\mathcal V_{7,E}^6}{r_{4,E}^2}.
\end{equation}
The factor $\mathcal V_{7,E}^6$ comes from the three inverse external metrics in the contraction, while the minus sign comes from the timelike index. Including the reduced measure, the electric contribution to the RR action becomes
\begin{equation}
\begin{aligned}
 S_{\mathrm{el}}
 &=-\frac1{4\kappa_{10}^2}
 \int d^{10}x\sqrt{-g_{10,E}}\,
 e^{\phi/2}|F_4^{\mathrm{el}}|_{10,E}^2\\
 &=-\frac1{4\kappa_3^2}
 \int d^3x\sqrt{-g_{3,E}}\,
 \frac{e^{\phi/2}}{\mathcal V_{7,E}^2}
 |F_4^{\mathrm{el}}|_{10,E}^2\\
 &=\frac1{\kappa_3^2}
 \int d^3x\sqrt{-g_{3,E}}\,
 \frac14\mathcal K_{\mathrm{el}}\mathfrak f^2,
 \qquad
 \mathcal K_{\mathrm{el}}
 =e^{\phi/2}\frac{\mathcal V_{7,E}^4}{r_{4,E}^2}.
\end{aligned}
 \label{eq:dr-electric-coefficient}
\end{equation}

To complete the electric contribution to the scalar potential, we still have to dimensionally reduce the topological terms involving $\mathcal D_2$ and then integrate out this field at fixed flux. For massless type IIA, the relevant topological action can be written in our conventions as
\begin{equation}
\begin{aligned}
 S_{\mathrm{top}}
 =\frac1{4\kappa_{10}^2}\int
 \bigg[
 &B_2\wedge F_4^{(10)}\wedge F_4^{(10)}
 -B_2\wedge B_2\wedge F_2\wedge F_4^{(10)}\\
 &
 +\frac13B_2\wedge B_2\wedge B_2\wedge F_2\wedge F_2
 \bigg],
 \label{eq:dr-topological-action}
 \end{aligned}
\end{equation}
as follows from the massless topological polynomial in \cite[eq.~(2.23)]{Bergshoeff:2001pv}. Its overall sign is consistent with the duality relation \eqref{eq:dr-duality} and the field equation $dF_6-H_3\wedge F_4^{(10)}=0$. Since $F_4^{\mathrm{el}}=\mathcal H_3\wedge\vartheta$ already contains three external legs, it can contribute only linearly to \eqref{eq:dr-topological-action}. Moreover, its product with the background four-form $F_4^{\mathrm{bg}}$ vanishes because both contain the circle direction.
The terms involving $\mathcal D_2$ therefore reduce to
\begin{equation}
 S_{\mathrm{top}}\big|_{\mathcal D_2}
 =\frac1{4\kappa_{10}^2}\int
 \mathcal H_3\wedge\vartheta\wedge
 \left(2B_2\wedge\mathcal G_4
       -B_2\wedge B_2\wedge F_2\right),
       \label{eq:StopD2}
\end{equation}
where the factor of two comes from the two occurrences of $F_4^{(10)}$ in the first term of \eqref{eq:dr-topological-action}.
The cubic term in $B_2$ has no dependence on $\mathcal D_2$
and vanishes in this ansatz because all its legs lie in
the six-dimensional factor.
Using the expansions of $B_2$, $F_2$ and $\mathcal G_4$, together with \eqref{eq:dr-basis-identities}, we obtain
\begin{align}
 B_2\wedge\mathcal G_4
 &=-\left(
 b_1\mathcal R_{23}+b_2\mathcal R_{13}
 +b_3\mathcal R_{12}\right)
 \widehat{\mathrm{vol}}_6,
 \nonumber\\
 B_2\wedge B_2\wedge F_2
 &=-2\left(
 m_0b_2b_3+Nb_1b_3+R_Fb_1b_2
 \right)\widehat{\mathrm{vol}}_6.
\end{align}
Substituting \eqref{eq:dr-axion-polynomials} then gives
\begin{equation}
 2B_2\wedge\mathcal G_4-B_2\wedge B_2\wedge F_2
 =2\rho\,\widehat{\mathrm{vol}}_6,
 \qquad
 \rho=-hb_1\Xi-m_0b_2b_3-Nb_1b_3-R_Fb_1b_2.
 \label{eq:dr-electric-polynomial}
\end{equation}
 Using $\mathcal H_3\wedge\mathcal A=0$, integration of \eqref{eq:StopD2} over the internal space yields
\begin{equation}
 S_{\mathrm{top}}\big|_{\mathcal D_2}
 =\frac1{2\kappa_3^2}\int\rho\,\mathcal H_3
 =\frac1{\kappa_3^2}\int d^3x\sqrt{-g_{3,E}}\,
 \frac12\rho\mathfrak f.
\end{equation}
Combining this with \eqref{eq:dr-electric-coefficient}, the complete $\mathcal D_2$-dependent action is
\begin{equation}
 S_{\mathcal D_2}
 =\frac1{\kappa_3^2}\int d^3x\sqrt{-g_{3,E}}\,
 \left(
 \frac14\mathcal K_{\mathrm{el}}\mathfrak f^2
 +\frac12\rho\mathfrak f
 \right).
 \label{eq:dr-topological-twoform}
\end{equation}

The equation for $\mathcal D_2$ is $d(\mathcal K_{\mathrm{el}}\mathfrak f+\rho)=0$, so $\mathcal K_{\mathrm{el}}\mathfrak f=-(M+\rho)$. Using \eqref{eq:dr-duality}, this gives
\begin{equation}
 (F_6)_{\mathrm{int}}
 =(M+\rho)\widehat{\mathrm{vol}}_6.
 \label{eq:dr-axion-F6}
\end{equation}
The integration constant $M$ is the Page flux and agrees with \eqref{RR6_expansion} at zero axions; indeed, $B_2\wedge\mathcal Q_3=0$ and \eqref{eq:dr-electric-polynomial} imply
\begin{equation}
 \left[
 F_6-B_2\wedge F_4^{(10)}
 +\frac12B_2\wedge B_2\wedge F_2
 \right]_{\mathrm{int}}
 =(M+\rho-\rho)\widehat{\mathrm{vol}}_6 = M \widehat{\mathrm{vol}}_6 .
\end{equation}
Eliminating $\mathcal D_2$ at fixed $M$ requires the boundary term $(2\kappa_3^2)^{-1}\int M\,d\mathcal D_2$ to ensure the correct sign of the scalar potential contribution. Including it and substituting the electric solution gives
\begin{equation}
 \left[
 \frac14\mathcal K_{\mathrm{el}}\mathfrak f^2
 +\frac12(M+\rho)\mathfrak f
 \right]_{\mathfrak f=-(M+\rho)/\mathcal K_{\mathrm{el}}}
 =-\frac{(M+\rho)^2}{4\mathcal K_{\mathrm{el}}}.
\end{equation}
The resulting contribution to the potential is therefore
\begin{equation}
 V_{\mathrm{el}}
 =\frac{(M+\rho)^2}{4\mathcal K_{\mathrm{el}}}
 =\frac{e^{-\phi/2}r_{4,E}^2}{4\mathcal V_{7,E}^4}
   (M+\rho)^2
 =\frac{e^{4D}g_s^2}{4\mathcal V_{6,S}^2}(M+\rho)^2.
 \label{eq:dr-electric-axion-potential}
\end{equation}
where we used \eqref{eq:dr-frame-dictionary},
\eqref{eq:dr-volumes} and \eqref{eq:dr-string-metric}.
At zero axions, it reduces to
\begin{equation}
 V_{F_6} =\frac{e^{4D}g_s^2M^2}{4\mathcal V_{6,S}^2}.
 \label{eq:dr-F6E}
\end{equation}

\paragraph{O6 tensions}

For a source wrapping $\Sigma_I$, the induced external measure
contributes $\mathcal V_{7,E}^{-3}$ and the internal integral gives
the wrapped volume $\mathrm{Vol}_E(\Sigma_I)$.
Reducing \eqref{eq:dr-source-action} and using
$2\kappa_{10}^2T_I=Q_{\mathrm{O6}}=-hm_0$ therefore gives
\begin{equation}
 V_{\mathrm{O6}}
 =\frac{e^{3\phi/4}Q_{\mathrm{O6}}}{2\mathcal V_{7,E}^3}
   \sum_I\mathrm{Vol}_E(\Sigma_I).
\end{equation}
For the cycles $2345$, $1346$, $1247$ and $4567$, dividing the
wrapped volume by $\mathcal V_{7,E}$ gives the inverse volume of
the respective transverse directions, hence
\begin{equation}
\begin{split}
 V_{\mathrm{O6}}
 &=-\frac{e^{3\phi/4}hm_0}{2\mathcal V_{7,E}^2}
 \left(\frac1{r_{1,E}r_{6,E}r_{7,E}}
       +\frac1{r_{2,E}r_{5,E}r_{7,E}}
       +\frac1{r_{3,E}r_{5,E}r_{6,E}}
       +\frac1{r_{1,E}r_{2,E}r_{3,E}}\right)\\
 &=-\frac{e^{4D}g_s hm_0}{2}
 \left(\frac1{r_1r_6r_7}
       +\frac1{r_2r_5r_7}
       +\frac1{r_3r_5r_6}
       +\frac1{r_1r_2r_3}\right)\\
 &=-\frac{e^{4D}g_s^2hm_0}{2\mathcal V_{6,S}}
 \sum_Ax_A.
\end{split}
 \label{eq:dr-sourceE}
\end{equation}
Each $\omega_a$ has at least one leg transverse to every wrapped cycle, so $P_I[\omega_a]=0$ and hence $P_I[B_2]=0$. Therefore, the $B_2$ axions we considered above do not contribute to the O6-plane terms.

\subsubsection*{Total scalar potential}

Collecting the curvature contribution \eqref{eq:dr-curvature},
the RR flux contributions \eqref{eq:dr-F2E},
\eqref{eq:dr-F4E} and \eqref{eq:dr-F6E}, and the O6
tensions \eqref{eq:dr-sourceE}, the potential at zero axions is
\begin{equation}
\begin{split}
 V_{\mathrm{no\,axions}}
 =e^{4D}\Bigg[&
 \frac14\sum_{i=1}^4a_i^2
 +\frac{g_s^2}{4}\left(
 \frac{m_0^2}{t_1^2}
 +\frac{N^2}{t_2^2}
 +\frac{R_F^2}{t_3^2}
 +|F_4^{\mathrm{bg}}|_{7,S}^2
 +\frac{M^2}{\mathcal V_{6,S}^2}\right)\\
 &-\frac{g_s^2hm_0}{2\mathcal V_{6,S}}
 \sum_Ax_A\Bigg].
\end{split}
 \label{eq:dr-compact-potential}
\end{equation}
For nonzero axions, we include the internal field strengths
$H_{3,\mathrm{int}}$ and $\mathcal G_4$ in
\eqref{eq:dr-full-field-strengths}, whose coefficients are given
in \eqref{eq:dr-axion-polynomials}, and replace the six-form
flux contribution \eqref{eq:dr-F6E} by the electric potential
\eqref{eq:dr-electric-axion-potential}. The full potential is
therefore
\begin{equation}
 V_{\mathrm{full}}
 =V_{\mathrm{no\,axions}}
 +\frac{e^{4D}}4\left[
 |H_{3,\mathrm{int}}|_{7,S}^2
 +g_s^2|\mathcal G_4|_{7,S}^2
 +\frac{g_s^2}{\mathcal V_{6,S}^2}
 \big((M+\rho)^2-M^2\big)
 \right],
 \label{eq:dr-full-axion-potential}
\end{equation}
where $\rho$ is defined in \eqref{eq:dr-electric-polynomial}; the subtraction $-M^2$ removes the zero-axion electric contribution already included in $V_{\mathrm{no\,axions}}$.
The complete potential is invariant under the gauged shifts \eqref{eq:dr-axion-gauging}: the RR axions enter only through $\Xi=\sum_A\xi_A$, for which $\delta\Xi=\lambda\sum_AQ_A=0$, while the NS axions $b_a$ are neutral.

As a check, we evaluate the potential on the general solution \eqref{eq:flux-radius-relations}, at zero axions and $D=D_0$. The four-form Bianchi identity \eqref{eq:dF4tad}, evaluated using \eqref{eq:flux-radius-relations}, gives $\sum_i a_i^2=\Delta^2/3$. Substituting the expressions for $m_0,N,R_F,f_i$ and $M$ in \eqref{eq:flux-radius-relations} into \eqref{eq:dr-F2E}, \eqref{eq:dr-F4E} and \eqref{eq:dr-F6E}, and using $\Delta=a_1-a_3+a_2-a_4$ together with $\sum_i a_i^2=\Delta^2/3$, gives the RR contributions below. For the source term \eqref{eq:dr-sourceE}, we also use the identity 
\begin{equation}
 h t_1\left(
 \frac1{r_1r_6r_7}
 +\frac1{r_2r_5r_7}
 +\frac1{r_3r_5r_6}
 +\frac1{r_1r_2r_3}
 \right)=\Delta.
\end{equation}
Together, these give
\begin{equation}
\begin{aligned}
 \sum_{i=1}^4a_i^2&=\frac{\Delta^2}{3},
 &g_s^2|F_2|_{7,S}^2&=\frac{2\Delta^2}{3},\\
 g_s^2|F_4^{\mathrm{bg}}|_{7,S}^2&=\frac{\Delta^2}{9},
 &\frac{g_s^2M^2}{\mathcal V_{6,S}^2}&=\frac{\Delta^2}{9},\\
 e^{-4D_0}V_{\mathrm{O6}}&=-\frac{\Delta^2}{3}.
\end{aligned}
\end{equation}
Consequently,
\begin{equation}
 V_0=e^{4D_0}\Delta^2
 \left(\frac1{12}+\frac16+\frac1{36}+\frac1{36}-\frac13\right)
 =-\frac{e^{4D_0}\Delta^2}{36}
 =-\frac1{L_{3,E}^2},
\end{equation}
which agrees with the Einstein equation in the normalization $R_{3,E}/2-V$, using $L_{\mathrm{AdS}}=6/|\Delta|$ and \eqref{eq:dr-AdS-radius}.

\subsection{First-order formulation}
\label{app:dr-CS}

The direct reduction gives the Maxwell action \eqref{eq:dr-3d-action}. To connect it with the formulation used in section~\ref{sec:3d-effective-theory}, we rewrite the couplings of $\mathcal{F}_2=d\mathcal{A}$ in an equivalent first-order form \cite{Nicolai:2003bp}, introducing the real scalar $\upsilon$ completing the complex scalar $Y=y+i\upsilon$. This rewriting does not add physical degrees of freedom and the other terms in the effective actions are unchanged.

We introduce an auxiliary one-form $\mathcal B$ and define
\begin{equation}
 D\upsilon=d\upsilon-\mathcal B,
 \qquad
 \delta\upsilon=\widetilde\lambda,
 \qquad
 \delta\mathcal B=d\widetilde\lambda,
\end{equation}
and we then replace the Maxwell term \eqref{eq:dr-KK-Maxwell} by
\begin{equation}
\begin{split}
 S_{\mathrm{first}}
 ={}&-\frac1{\kappa_3^2}
 \int d^3x\sqrt{-g_{3,E}}\,
 \frac{1}{4y^2}D_\mu\upsilon D^\mu\upsilon+\frac1{2\kappa_3^2}
 \int\mathcal B\wedge d\mathcal A.
\end{split}
 \label{eq:dr-first-order}
\end{equation}
The second term is a BF coupling, invariant under the shift of $\mathcal B$ up to a boundary term. 
To verify the equivalence between the two formulations, set
$\widetilde{\mathcal F}^{\mu}=\frac12\varepsilon^{\mu\nu\rho}\mathcal F_{\nu\rho}$, where $\varepsilon^{\mu\nu\rho}$ is the Levi--Civita tensor
of $g_{3,E}$, and vary \eqref{eq:dr-first-order} with respect to $\mathcal B$, obtaining
\begin{equation}
 D_\mu \upsilon=\partial_\mu\upsilon-\mathcal B_\mu  =-y^2\widetilde{\mathcal F}_\mu.
\end{equation}
Substituting back into \eqref{eq:dr-first-order} yields
\begin{equation}
 S_{\mathrm{first}}\big|_{\mathcal B\ \mathrm{eliminated}}
 =
 -\frac1{\kappa_3^2}
 \int d^3x\sqrt{-g_{3,E}}\,
 \frac{y^2}{8}\mathcal F_{\mu\nu}\mathcal F^{\mu\nu}
 +\frac1{2\kappa_3^2}\int d\upsilon\wedge d\mathcal A,
\end{equation}
where we used $\widetilde{\mathcal F}_\mu\widetilde{\mathcal F}^{\mu} =-\frac12\mathcal F_{\mu\nu}\mathcal F^{\mu\nu}$ in mostly-plus signature. Since the last term is a total derivative, this reproduces \eqref{eq:dr-KK-Maxwell}.

After eliminating the external two-form at fixed Page flux,
the complete first-order action is therefore
\begin{equation}
\begin{split}
 S_{3D}^{\mathrm{first}}
 ={}&\frac1{\kappa_3^2}
 \int d^3x\sqrt{-g_{3,E}}
 \left[
 \frac12R_{3,E}
 -\mathcal L_{\mathrm{scal}}
 -\frac{1}{4y^2} D_\mu\upsilon D^\mu\upsilon 
 -V_{\mathrm{full}}
 \right]+\frac1{2\kappa_3^2}
 \int\mathcal B\wedge d\mathcal A,
\end{split}
 \label{eq:dr-complete-retained-bosonic-action}
\end{equation}
with $\mathcal L_{\mathrm{scal}}$ and $V_{\mathrm{full}}$
given in \eqref{eq:dr-Maxwell-description} and
\eqref{eq:dr-full-axion-potential}, respectively.
The kinetic terms for $y$ (contained in $\mathcal{L}_{\mathrm{scal}}$) and $\upsilon$ combine into $D_\mu YD^\mu\bar Y/(4y^2)$, and the complex scalar coordinates have covariant derivatives
\begin{equation}
 DY=dY-i\mathcal B,\qquad
 DX^A=dX^A-iQ_A\mathcal A,\qquad
 DT_a=dT_a.
 \label{eq:dr-final-covariant-derivatives}
\end{equation}
In particular, $\mathcal A$ continues to gauge the RR axion shifts determined by \eqref{eq:dr-axion-gauging}. The interpretation in terms of gauged supergravity is discussed in section~\ref{sec:3d-effective-theory}.

\bibliographystyle{utphys}
\bibliography{refs}

\end{document}